\documentclass{aa}  
\usepackage{natbib}
\usepackage{graphicx}
\usepackage{footmisc}
\usepackage{float}
\usepackage{caption}
\usepackage[caption=false]{subfig}
\usepackage[colorlinks=true,citecolor=blue]{hyperref}
\usepackage{multirow}
\usepackage{amsmath}
\usepackage{nccmath}
\usepackage{color}
\usepackage{array} 
\usepackage[flushleft]{threeparttable}
\DeclareTextFontCommand{\textroman}{\fontlibertine}
\usepackage{txfonts}

\begin{document}

   \title{Detection of sodium in the atmosphere of WASP-69b}


   \author{N. Casasayas-Barris\inst{1,2} \and E. Palle\inst{1,2} \and G. Nowak\inst{1,2} \and F. Yan\inst{3} \and L. Nortmann\inst{1,2} \and F. Murgas\inst{1,2}}

   \institute{Instituto de Astrofísica de Canarias, Vía Láctea s/n, E-38205 La Laguna, Tenerife, Spain
              \\
              \email{nuriacb@iac.es}
         \and
             Departamento de Astrofísica, Universidad de La Laguna, Spain
        \and
            Max Planck Institute for Astronomy, Königstuhl 17, 69117 Heidelberg, Germany}
             

   \date{Received Month 00, 2017; accepted Month 00, 2017}

 
  \abstract
   {Transit spectroscopy is one of the most commonly used methods to characterize exoplanets' atmospheres. From the ground, these observations are very challenging due to the terrestrial atmosphere and its intrinsic variations, but high-spectral resolution observations overcome this difficulty by resolving the spectral lines and taking advantage of the different Doppler velocities of the Earth, the host star and the exoplanet. 
   }
   {We analyze the transmission spectrum around the Na I doublet at $589~\mathrm{nm}$ of the extrasolar planet WASP-69b, a hot Jupiter orbiting a K-type star with a period of $3.868$ days, and compare the analysis to that of the well-know hot Jupiter HD~189733b. We also present the analysis of the Rossiter-McLaughlin effect for WASP-69b.}
   {We observed two transits of WASP-69b with the High Accuracy Radial velocity Planet Searcher (HARPS-North)  spectrograph ($R=115~000$) at the TNG telescope. We perform a telluric contamination subtraction based on the comparison between the observed spectra and a telluric water model. Then, the common steps of the differential spectroscopy are followed to extract the transmission spectrum. The method is tested with archival transit data of the  extensively studied exoplanet HD~189733b, obtained with the HARPS-South spectrograph at ESO $3.6~\mathrm{m}$ telescope, and then applied to WASP-69b data.}
   {For HD~189733b, we spectrally resolve the Na I doublet and measure line contrasts of $0.72\pm0.05~\%$ (D2) and $0.51\pm0.05~\%$ (D1), and FWHMs of $0.64\pm0.04~\mathrm{\AA}$ (D2) and $0.60\pm0.06~\mathrm{\AA}$ (D1), in agreement with previously published results. For WASP-69b only the contrast of the D2 line can be measured ($5.8\pm0.3~\%$). This corresponds to a detection at the $5\sigma$-level of excess absorption of $0.5\pm0.1~\%$ in a passband of $1.5~\mathrm{\AA}$. A net blueshift of ${\sim}0.04~\mathrm{\AA}$ is measured for HD~189733b and no shift is obtained for WASP-69b. By measuring the Rossiter-McLaughlin (RM) effect, we get an angular rotation of $0.24^{+0.02}_{-0.01}~\mathrm{rad/day}$ and a sky-projected angle between the stellar rotation axis and the normal of orbit plane ($\lambda$) of $0.4^{+2.0}_{-1.9}~^o$ for WASP-69b. Similar results to those previously presented in the literature are obtained for the RM analysis of HD~189733b.}
   {Even if sodium features are clearly detected in the WASP-69b transmission spectrum, more transits are needed to fully characterize the lines profiles and retrieve accurate atmospheric properties.}

   \keywords{Planetary systems -- Planets and satellites: individual: WASP-69b -- Planets and satellites: individual: HD189733b -- Planets and satellites: atmospheres -- Methods: observational -- Techniques:  spectroscopic               }

   \maketitle
%

\section{Introduction}

Over the last two decades, there has been an enormous progress in the search for planets outside our solar system, resulting in thousands of exoplanet detections. Moreover, the atmospheric characterization of some of these discovered planets has been rapidly expanding with the constantly improvement of the astrophysical instrumentation, being the transiting systems the most amenable targets for atmospheric studies. Transmission spectroscopy is one of the best known methods for atmospheric characterization. During a transit, the stellar light penetrates the planetary atmosphere and its signatures appear imprinted in the stellar flux. Observing when the planet is crossing the stellar disk and when it is not, the transmission spectrum of the planet can be extracted through differential spectroscopy, comparing the in- and out-of-transit measurements. 

The first detection of an exoplanet atmosphere by  \citealt{2002ApJ...568..377C}, which revealed the presence of atomic sodium (Na I), was only possible thanks to space-based instruments on-board the Hubble Space Telescope (HST). For a time, ground-based observations were thought to be too challenging, but the difficulties due to the terrestrial atmospheric variations and systematic effects during the observations of transiting systems have been slowly overcome, resulting in robust detections of Na, K and Rayleigh-like slopes with low spectral resolution spectrographs (\citealt{2012Sing}, \citealt{2014Murgas}, \citealt{2015Wilson},  \citealt{2017Chena}, \citealt{2017Palle}, among others). The larger aperture of ground-based telescopes provides a significant advantage over space-based observations, but limitations still remain due to telluric absorption in the atmosphere. However, observations at high-spectral resolution overcome this difficulty of dealing with the telluric atmosphere by resolving the spectral lines and taking advantage of the different Doppler velocities of the Earth, the host star and the exoplanet. 

The Na I doublet, with lines at $5895.924~\mathrm{\AA}$ (D1) and $5889.951~\mathrm{\AA}$ (D2), is one of the easiest species to detect in a hot planet upper atmosphere due to its high cross-section, and the line cores can trace the temperature profiles up to the planet thermosphere. The first ground-based detection of Na I doublet in HD~189733b  was performed with the High Resolution Spectrograph (HRS), with $R{{\sim}}60~000$, mounted on the 9.2$~\mathrm{m}$ Hobby-Eberly Telescope \citep{2008ApJ...673L..87R}. Soon after, the Na I was confirmed in HD~209458b using the High Dispersion Spectrograph (HDS), with $R{\sim}45~000$, on the $8~\mathrm{m}$ Subaru Telescope \citep{2008Snellen}. Recent ground-based studies with the High Accuracy Radial velocity Planet Searcher (HARPS) spectrograph ($R{\sim}115~000$), at ESO $3.6~\mathrm{m}$ telescope in La Silla (Chile), have been able to resolve the individual Na I line profiles of HD~189733b \citep{2015A&A...577A..62W} and WASP-49b \citep{2017A&A...602A..36W}. Several studies have used these same data sets, e.g., \citealt{2015Heng} studied the existence of temperature gradients, \citealt{2015ApJ...814L..24L} explored high-altitude winds in the atmosphere of HD 189733b, \citealt{2016MNRAS.462.1012B} studied stellar activity signals, and \citealt{2017A&A...603A..73Y} focused on center-to-limb variation effect (CLV) in the transmission light curve of this same exoplanet.

High resolution transmission spectroscopy methods will be crucial in the near-future in order to observe spectral features of rocky planets and super-Earths, which will be out of reach of the James Webb Space Telescope (JWST), but available with the upcoming facilities such as ESPRESSO on the Very Large Telescope (VLT) or HIRES on E-ELT (\citealt{2013Snellen}, \citealt{2017Lovis}). 

Here, we present the results obtained with the HARPS-North spectrograph by analyzing two transit observations of WASP-69b. WASP-69b \citep{2014MNRAS.445.1114A} is a Saturn-mass planet ($0.26M_{Jup}$, $1.06R_{Jup}$) in a $3.868~\mathrm{days}$ orbit around a K5 star. Its large atmospheric scale height (${\sim}650~\mathrm{km}$) and the small size of the star make this planet a good target for transmission spectroscopy (see Table~\ref{table:planet_param}). We present an alternative reduction method, which we apply firstly to the well-studied exoplanet HD~189733b, using data from previous HARPS observations, and then to the new WASP-69b observations. 

This paper is organized as follows. In Sect. 2, we present the observations. In Sect. 3, we describe the data reduction process, including the telluric correction and the transmission spectrum extraction. In Sect. 4, we present and discuss the results obtained on the atomic sodium (Na I) absorption observed in the transmission spectrum and the transmission light curve of both HD~189733b and WASP-69b. In Sect. 5, we present the results of the Rossiter-McLaughlin effect measurement of WASP-69b.

\renewcommand{\thefootnote}{\fnsymbol{footnote}}

\begin{table*}[h]
\centering
\caption{Physical and orbital parameters of WASP-69 and HD~189733 systems.}
\begin{tabular}{llrr}
\hline\hline
 & Parameters   & \multicolumn{2}{c}{Values}\\
\cline{3-4}
          & &WASP-69 & HD~189733 \\
\hline
&$V$ mag            & $9.9$& $7.7$\\
&Spectral type      & K5& K1-K2\\
&$T_{eff}$ [K]      & $4700\pm50$&$5040\pm50$\footnote[1]{}\\
\textbf{Star}&$V~\sin{i}$ [km/s]      & $2.20\pm0.4$ & $3.5\pm1$\footnote[1]{}\\
&$M_{\star}$ [$M_{\odot}$] &$0.826\pm0.029$&$0.806\pm0.048$\footnote[1]{}\\
&$R_{\star}$ [$R_{\odot}$] &$0.813\pm0.028$&$0.756\pm0.018$\footnote[1]{}\\
\hline
&$T_{0}$ [BJD]      & $2455748.83344\pm0.00018$ &$2454279.436714\pm0.000015$\footnote[4]{}\\
&$P$ [days]         & $3.8681382\pm1.7x10^{-6}$&$2.21857567\pm1.5x10^{-7}$\footnote[4]{}\\
&$t_T$ [days]       & $0.0929\pm0.0012$ &$0.0760\pm0.0017$\footnote[9]{}\\
&$a$ [AU]           & $0.04525\pm0.00075$&$0.03100\pm0.00062$\footnote[2]{}\\
&$a/R_{\star}$      & $12.00\pm0.46$&$8.84\pm0.27$\footnote[4]{}\\
&$b$ [$R_{\star}$]  & $0.686\pm0.023$ & $0.6631\pm0.0023$\footnote[4]{}\\
\textbf{Transit} &$i_p$ [degrees]      & $86.71\pm0.2$&$85.7100\pm0.0023$\footnote[4]{}\\
&$e$                & $0$&$0$\footnote[5]{}\\
&$\omega$ [degrees] & $90$&$90$\footnote[5]{}\\
&$\gamma$ [m/s]     & $-9.62826\pm0.00023$&$-2.57\pm0.143$\footnote[3]{}\\
&$K_{1}$ [m/s]      & $38.1\pm2.4$&$205.0\pm6$\footnote[5]{}\\
\hline
&$M_{P}$ [$M_{J}$]  & $0.260\pm0.0185$&$1.144\pm0.056$\footnote[2]{}\\
\textbf{Planet}&$R_{P}$ [$R_{J}$]  & $1.057\pm0.017$&$1.138\pm0.027$\footnote[1]{}\\
&$T_{eq}$[K]        & $963\pm18$&$1191\pm20$\footnote[6]{}\\
&$H$ [km]           & ${\sim}650$&${\sim}200$\\
\lasthline
\end{tabular}
\\
\begin{tablenotes}
\item Notes. All WASP-69 parameters are taken from \citet{2014MNRAS.445.1114A}. $^($\footnotemark[1]$^)$\citet{2008ApJ...677.1324T}. $^($\footnotemark[4]$^)$\citet{2010ApJ...721.1861A}. $^($\footnotemark[9]$^)$\citet{2003ApJ...585.1038S}. $^($\footnotemark[2]$^)$\citet{2006ApJ...646..505B}. $^($\footnotemark[5]$^)$\citet{2005A&A...444L..15B}. $^($\footnotemark[3]$^)$\citet{2009A&A...495..959B}.
$^($\footnotemark[6]$^)$\citet{TeffHD189}. 
\end{tablenotes}
\label{table:planet_param}
\end{table*}

\renewcommand{\thefootnote}{\arabic{footnote}}

\section{Observations}
We observed two transits of WASP-69b on 4 June 2016 and 4 August 2016 using the High Accuracy Radial velocity Planet Searcher (HARPS-North) spectrograph mounted on the 3.58 m Telescopio Nazionale Galileo (TNG), located at Roque de los Muchachos Observatory (ORM). For the first transit the observations started at 01:25 UT and finished at 05:19 UT. The airmass variation was from $2.2$ to $1.2$. In the second night, we observed from 22:29 to 02:51 UT, and the airmass changed from $1.5$ to $1.2$.

For both transits, we observed continuously during the night, exposing before, during and after the transit in order to retrieve a good baseline and a high signal-to-noise out-of-transit spectrum, which is important for the data reduction process. Several transits are needed to reach enough signal-to-noise for the detection of the spectroscopic exoplanetary signatures. The observations were carried out using fiber A on the target and fiber B on the sky. Since WASP-69 has a magnitude of $9.87$ (V), $900~\rm{s}$ of exposure time were used, recording a total of 35 spectra, 16 of them forming our in transit sample. The remaining spectra are out of transit. 

There is one further transit dataset of WASP-69b in the HARPS archive (program: 089.C-0151(B), PI: Triaud) taken on 22 June 2012. However, telluric Na emission is observed during that night and the sky fiber B was not used, impeding telluric decontamination. Thus, we have not used these data in this paper. 

HD~189733 was observed with HARPS on the ESO $3.6~\rm{m}$ telescope in La Silla, Chile. Data were retrieved from the ESO archive, programs 072.C-0488(E), 079.C-0828(A) (PI: Mayor) and 079.C-0127(A) (PI: Lecavelier des Etangs). In total, HD~189733 was observed during 4 transits, being the last night affected by bad weather, and which is not used here. Same as the WASP-69 observations, the whole nights were dedicated to retrieve the major baseline possible, recording 99 spectra in total, 46 of them obtained during transit. In these observations, sky spectra on fiber B were gathered only for the second and third nights. These data have been extensively studied by other authors, e.g. \citet{2009A&A...506..377T}, \citet{2015A&A...577A..62W}, \citet{2015ApJ...814L..24L}, \citet{2015A&A...580A..84D}, \citet{2016MNRAS.462.1012B}, and \citet{2017A&A...603A..73Y}. The observing logs of the two data sets employed here are summarized in Table~\ref{table:obs_info}.

\section{Methods}

\subsection{Data reduction}
\begin{figure}[h]
\centering
\includegraphics[width=0.49\textwidth]{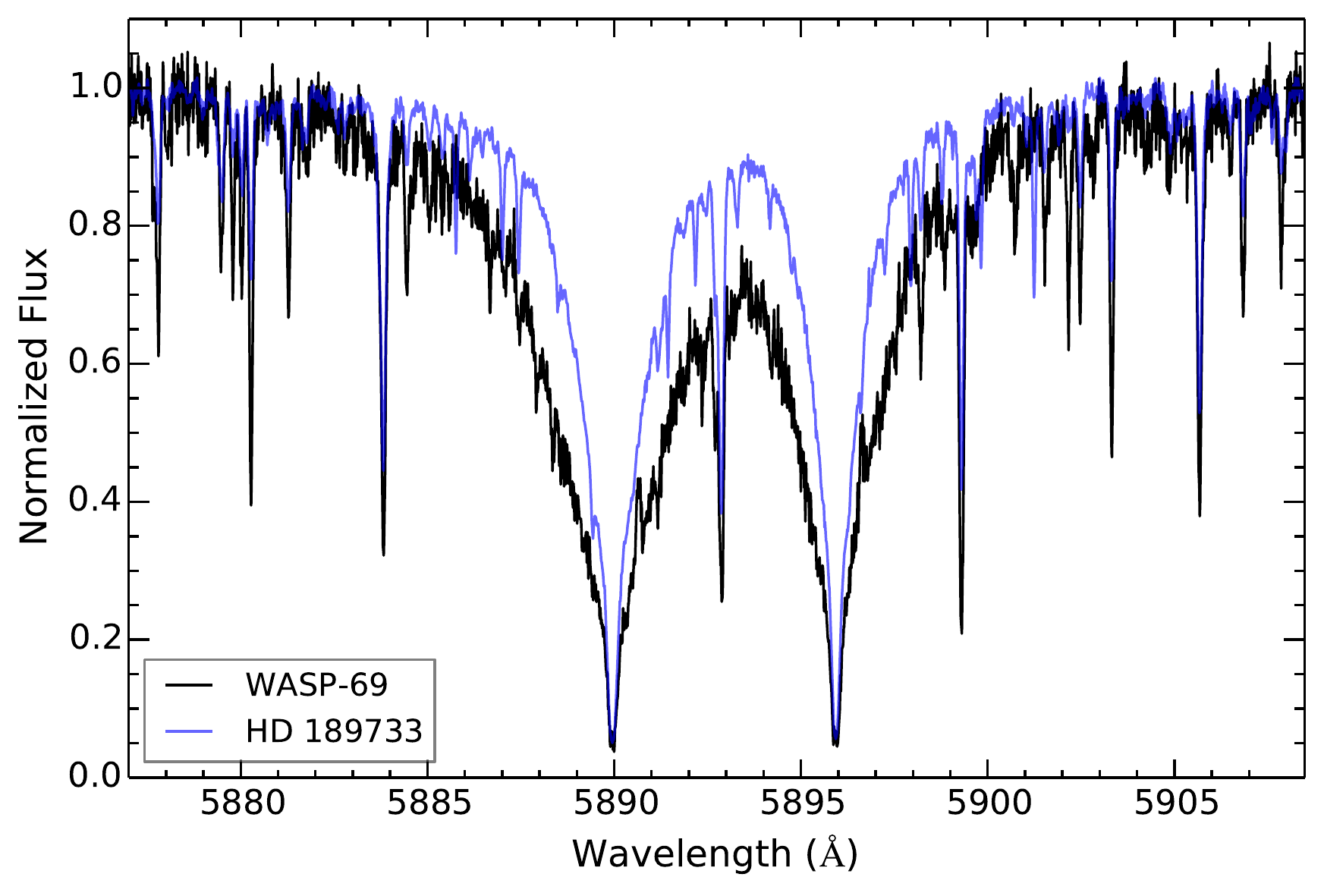}
\caption{Normalized spectra of WASP-69 (black) and HD~189733 (blue) after being reduced by the HARPS DRS around the Na I region. Both spectra are in their stellar reference frame.}
\label{fig:example_spec}
\end{figure}

The observations were reduced with the HARPS Data Reduction Software (DRS), version 1.1 of HARPS-North DRS for WASP-69 and version 3.5 of HARPS-South DRS for HD~189733. In both cases, the DRS extracts the spectra order-by-order, which are then flat-fielded using the daily calibration set. For each spectral order, a blaze correction is applied together with the wavelength calibration and, finally, all the spectral orders from each two-dimensional echelle spectrum are combined and resampled, ensuring flux conservation, into a one-dimensional spectrum. The resulting spectra are referred to the Solar System barycenter rest frame and the wavelengths are given in the air. The reduced one-dimensional spectra covers a wavelength range between $3800$ and $6900~\rm{\AA}$, with a wavelength step of $0.01~\rm{\AA}$ and  a spectral resolution of $R{\sim}115~000$. Two representative spectra, one for each planet, around the Na doublet region are shown in Figure~\ref{fig:example_spec}. 

\subsection{Telluric correction}

One of the major difficulties of ground-based observations is dealing with the telluric imprints from the Earth's atmosphere. Since the telluric transmission during a night depends on the airmass and on water column variations in the air, the atmospheric constituents contaminate the recorded spectra by producing a time-variable diversity of absorption and emission lines. The main contributors in the optical domain are water and molecular oxygen, nevertheless, we also expect telluric sodium signatures near the Na I doublet \citep{2008Snellen}. This telluric sodium suffers seasonal variations and, within a night, its behavior is different from the other telluric features, making it necessary to remove these two different contamination sources using different processes.

\subsubsection{Telluric sodium}

\begin{figure}[h]
\centering
\includegraphics[width=0.49\textwidth]{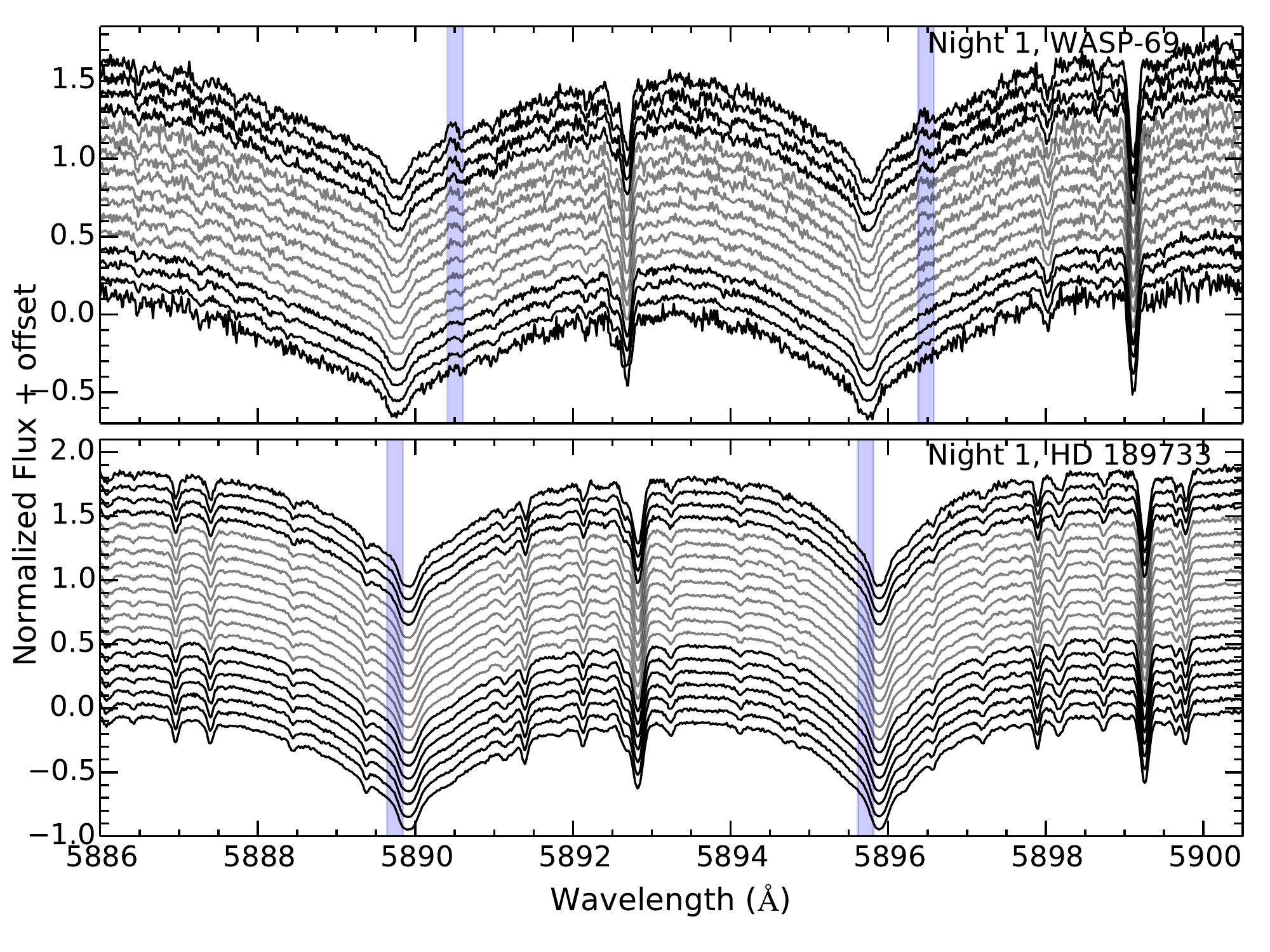}
\caption{Normalized spectra of WASP-69 (top) and HD~189733 (bottom) for Night 1, in the region of Na I doublet. Different flux offsets are added to each spectrum for a better visualization. In both cases, the spectra are organized by observing time: the first spectrum of the night is located in the upper position and the last one in the lowest position. For WASP-69 the airmass decreases with time and for HD~189733 it increases. In black are shown the out-of-transit spectra and in gray, the in-transit spectra. The blue regions show where the telluric sodium is observed for WASP-69 spectra and where the telluric sodium is expected for HD~189733 data. Note that the last spectrum of WASP-69 for the first night is not used in the reduction process because of the higher noise compared to other spectra of the same night.}
\label{fig:tell_Na}
\end{figure}

\begin{figure}
\centering
\includegraphics[width=0.49\textwidth]{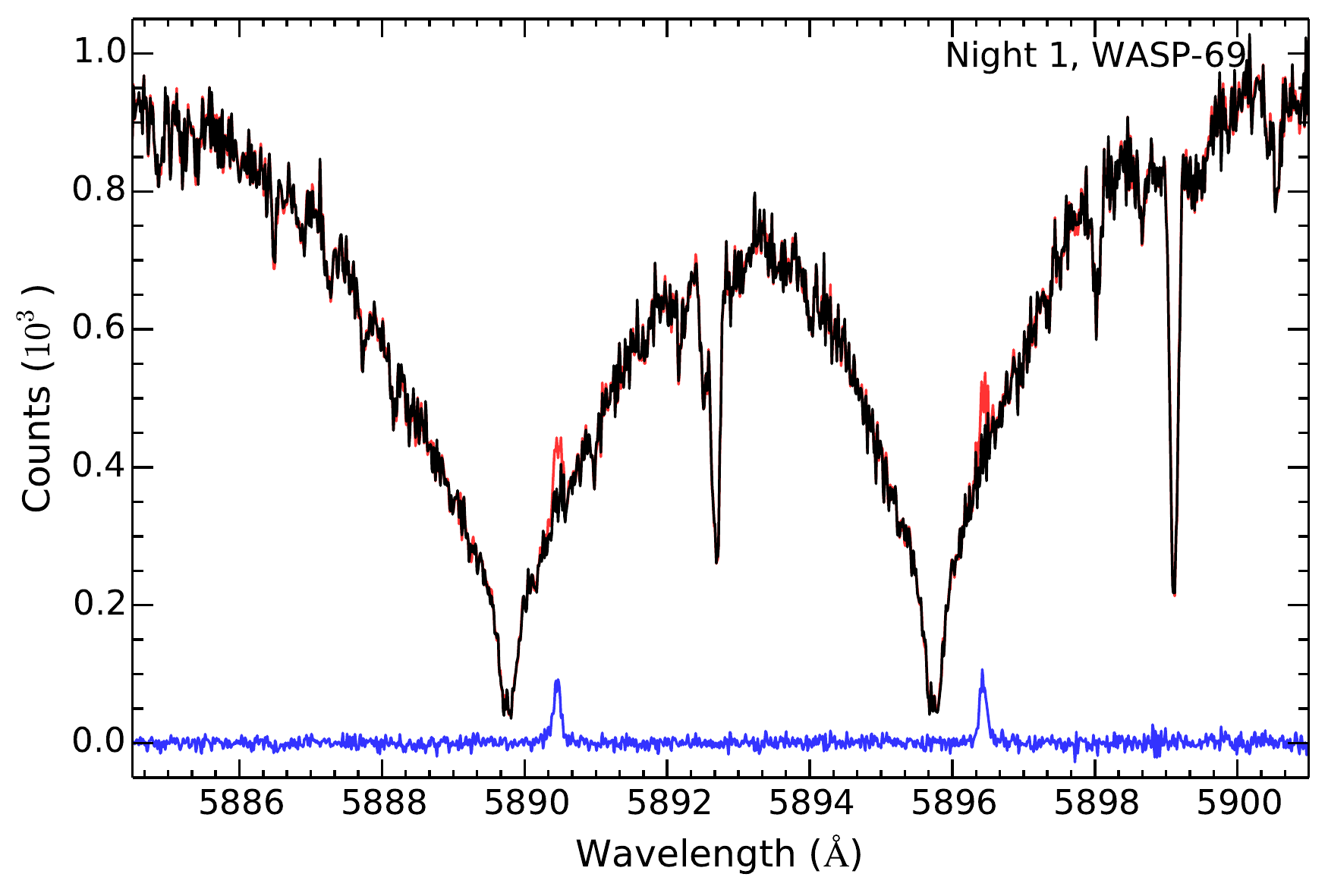}
\caption{Telluric sodium correction for WASP-69 spectra. The red spectrum is one of WASP-69 observed spectra for night 1, where the telluric sodium features can be  observed in the wings of the stellar Na I lines. The blue line is the corresponding sky spectrum taken with the HARPS-N fiber B, and the black line is the resulting WASP-69 spectrum after the sky subtraction.}
\label{fig:tell_na_subtrac}
\end{figure}

Each individual spectrum of both planets was examined for the presence of telluric sodium. 
For WASP-69b telluric sodium appears in the spectra as emission, as can be also observed in the sky spectra retrieved with fiber B simultaneously to the target data, which ensures exactly the same atmospheric conditions in both spectra (Figure \ref{fig:tell_Na}). Here, we remove the telluric sodium contamination in WASP-69 data by simply subtracting the sky spectra to the target spectra (see Figure \ref{fig:tell_na_subtrac}). The sky spectra taken together in HD~189733 observations were also checked but no sodium emission is observed.  

These telluric signatures present a strong dependence on airmass, being more intense at higher airmass and rapidly disappearing at lower airmass. However, its behavior is not exactly the same as the other telluric lines, producing strong telluric residuals in the resulting transmission spectrum if we make the assumption that they can be corrected as the other telluric features \citep{2015A&A...577A..62W}. This shows the importance of using fiber B monitoring the sky background simultaneously to the observations. In \citet{2017A&A...602A..36W} telluric sodium is observed in some of the WASP-49 spectra. They find a random behaviour of this telluric emission in their spectra, with no correlation between the sodium emission and the airmass. Not correcting for this effect could be a problem in case of a null barycentric Earth radial velocity (BERV) and a small stellar systemic velocity, where the telluric and exoplanetary sodium could overlap in the mid-transit position.

\subsubsection{Other telluric features}
In addition to removing telluric sodium, it is necessary to remove the telluric imprints produced by water vapor, the main contributor to the telluric contamination near the Na I doublet. 

\begin{figure}{}
\centering
\includegraphics[width=0.49\textwidth]{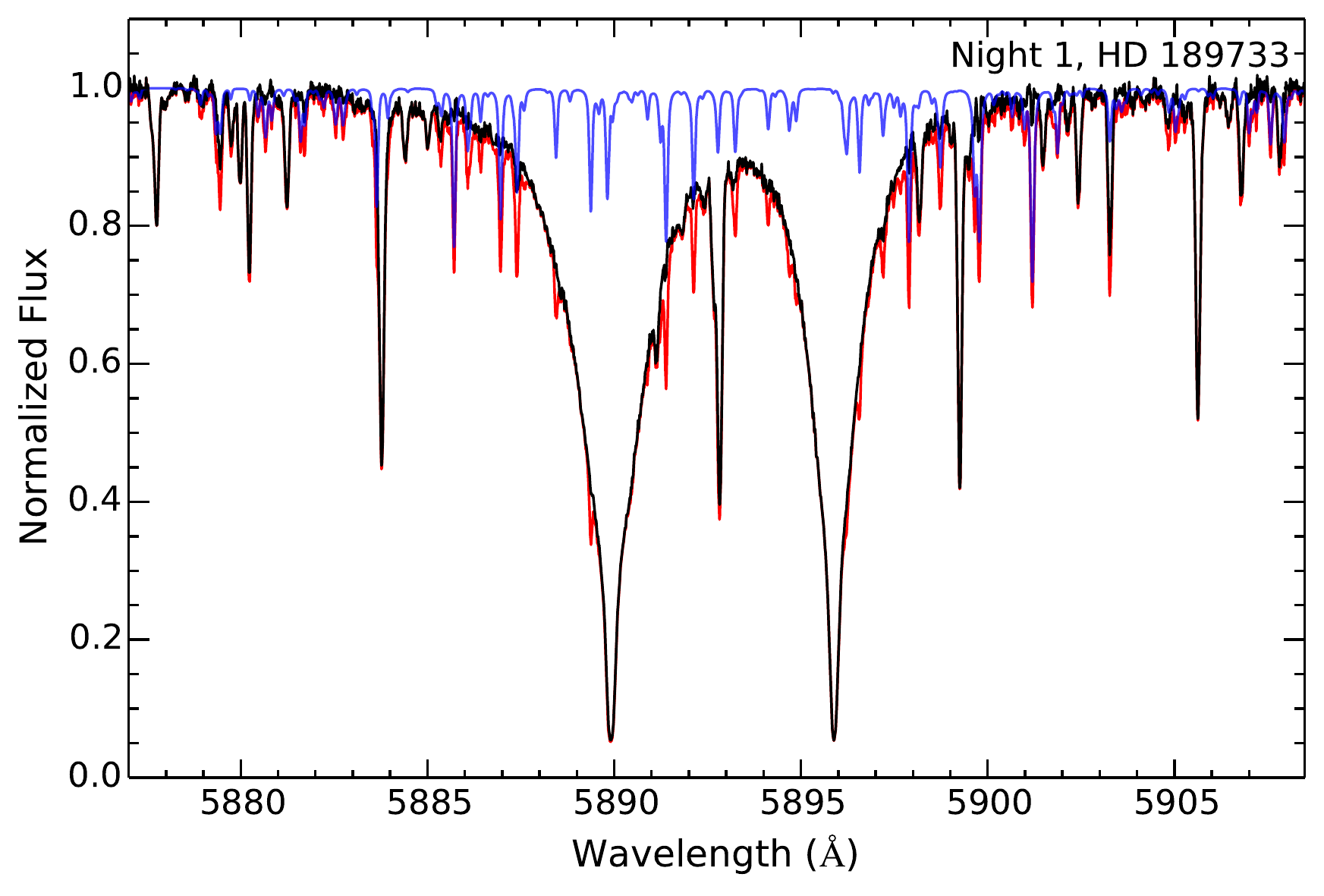}
\caption{Telluric correction of the observed spectra of HD~189733. The red spectrum is one of the observed HD~189733 spectra for night 1, the blue line is the shifted and scaled telluric water model from \citep{Yanmodel}, and the black spectrum is the resulting spectrum after applying the telluric correction.}
\label{fig:tell_corr}
\end{figure}

We present a different method to correct for the telluric contamination with respect to the methods described in \citet{2010A&A...523A..57V}, \citet{2013A&A...557A..56A} and \citet{2015A&A...577A..62W}, which consider that the variation of telluric lines follows the airmass variation linearly. Here, the telluric features are corrected by using the one-dimensional telluric water model presented in \citet{Yanmodel}, which uses the line list from HITRAN \citep{HITRAN}. The spectra are referred to the Solar System reference frame. In this frame the Earth's radial velocity variation during the night is reflected as a wavelength shift of the telluric lines with respect the air position. Note that the telluric lines absorption depth also change with the airmass. For each observed spectrum, we shift the telluric water vapor model to the position of the telluric lines using the barycentric Earth radial velocity information. This shifted model is then scaled to the same airmass of the observed spectrum and, finally, the observed spectrum is divided by the scaled telluric model, removing the telluric features from the data (see Figure \ref{fig:tell_corr}).   

\begin{table*}[h]
\centering
\caption{Observations log for WASP-69 and HD189733 for the employed data sets.}
\begin{tabular}{ccccccccc}
\hline\hline
& & Date & \# Spectra\footnote{} & Exp. Time [s] & Airmass & SNR\footnote{}& SNR\footnote{} & Fiber B\footnote{}\\
\hline
WASP-69 & Night 1 & 2016-06-04 & 8/16 & 900 & 2.2-1.2& ${\sim}45$& ${\sim}11$&Yes\\
& Night 2 & 2016-08-04 & 8/18 & 900 & 1.5-1.2& ${\sim}40$& ${\sim}8$&Yes\\
\hline
HD~189733 & Night 1 & 2006-09-07 & 9/20 & 900-600& 1.6-2.1& ${\sim}165$& ${\sim}40$&No\\
& Night 2 & 2007-07-19 & 18/39 & 300 & 2.4-1.6& ${\sim}115$& ${\sim}22$&Yes\\
& Night 3 & 2007-08-28 & 19/40 & 300& 2.2-1.6& ${\sim}100$& ${\sim}20$&Yes\\
\hline
\end{tabular}\\
\begin{tablenotes}
\item Notes. \footnotemark[1]Number of in-transit spectra / total number of spectra. \footnotemark[2]The signal-to-noise ratio (S/N) per pixel extracted in the continuum near 5900$~\rm{\AA}$. \footnotemark[3]S/N in the lines core of the Na I D lines. \footnotemark[4]Fiber B monitoring the sky simultaneously to the target observations.
\end{tablenotes}
\label{table:obs_info}
\end{table*}

The variation in the transmission of the Earth's atmosphere is one of the main problems when a model is used to correct the telluric contamination. The line depth variation is not equal for all the telluric lines, leading to poor results in the removal of the telluric signatures when scaling the model, and possibly introducing small residuals into the final transmission spectrum. However, one of the best points of this method is that no additional noise is introduced in the corrected spectra, key in transit spectroscopy, taking into account the low signal-to-noise expected for the atmospheric planetary signals. On the other hand, we also consider the telluric line's movement during the night, since omitting this shift could produce signals of misalignment in the next reduction steps. This alternative method can be also useful when the baseline of the observations is insufficient to compute a high-quality telluric spectrum for an efficient correction of telluric lines.

\subsection{Transmission spectrum}
Transit spectroscopy requires observations when the planet is crossing the stellar disk and when it is not. The out-of-transit spectra contain the stellar flux partially absorbed by the Earth's atmosphere, while the in-transit spectra contain, additionally, the exoplanet atmosphere transmission. These data are used to compute their ratio spectrum and measure the extra absorption caused by the planet atmosphere. 

\begin{figure}
\centering
\includegraphics[width=0.49\textwidth]{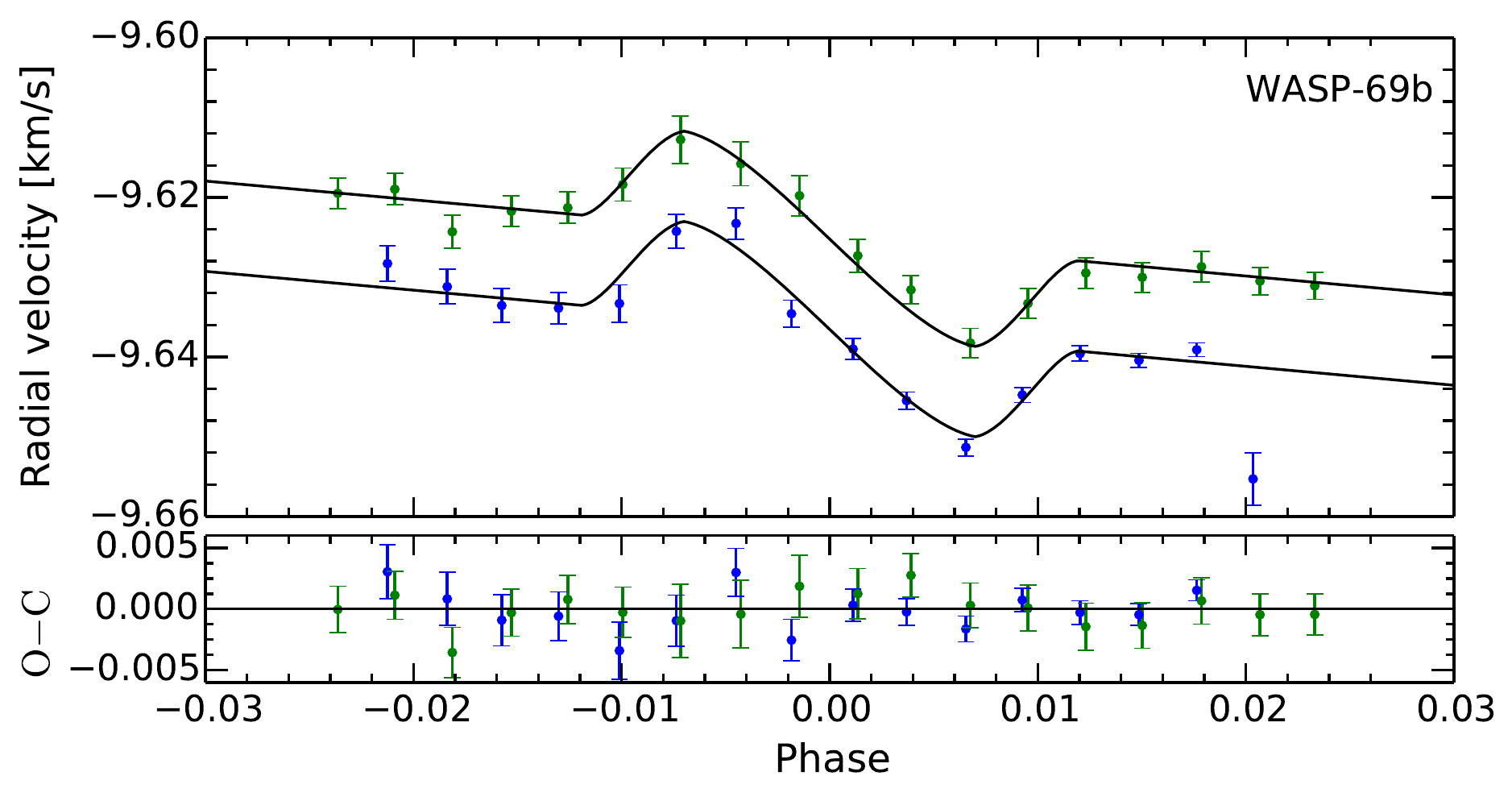}
\includegraphics[width=0.49\textwidth]{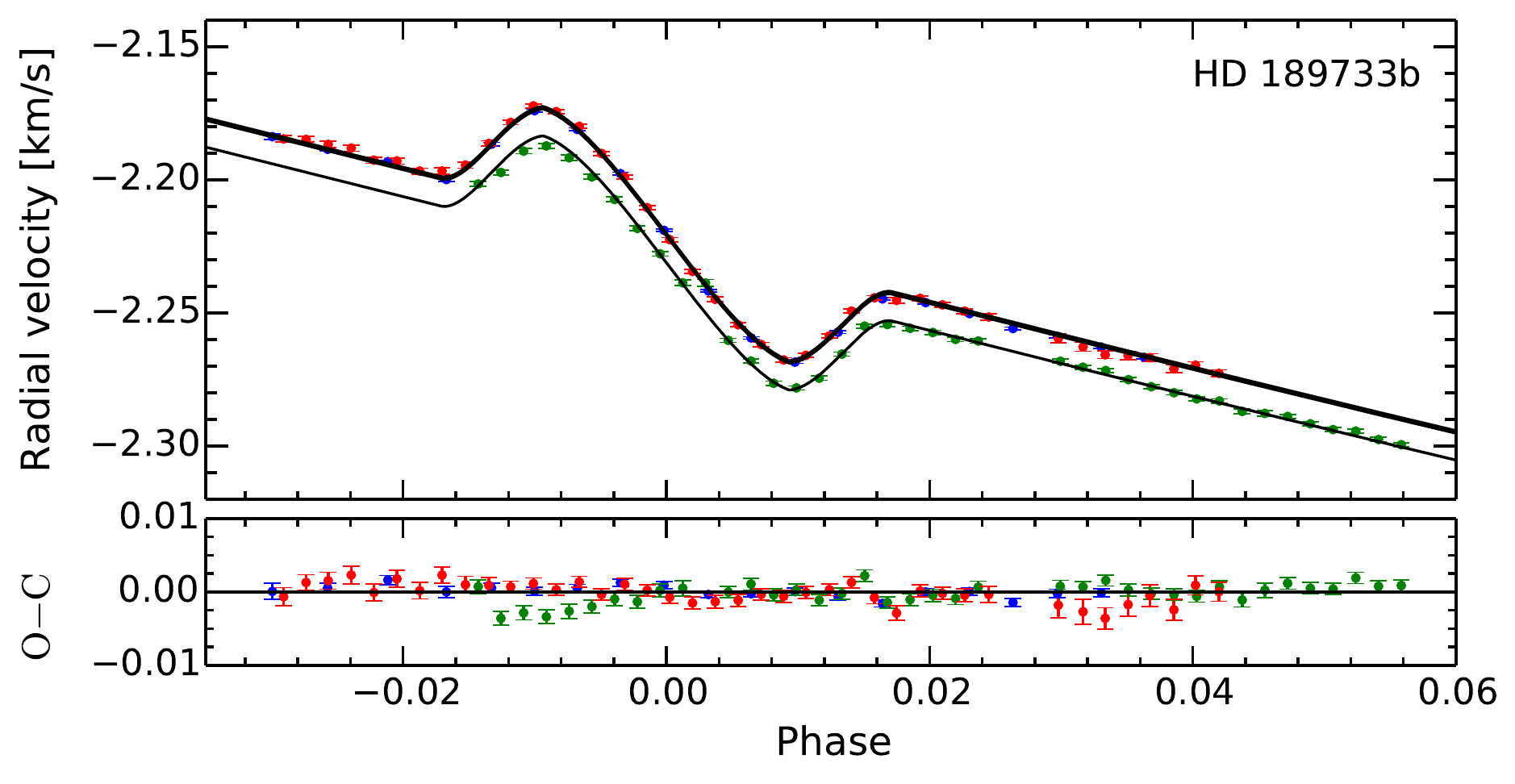}
\caption{Nightly evolution of the radial velocity of WASP-69 (top) and HD~189733 (bottom) (in blue the night 1, in green the night 2, and in red the night 3). In the vertical axis we show the radial velocity of the star in $\mathrm{km\cdot s^{-1}}$. The radial velocity and error values are taken from the HARPS files header. The black solid line is the best fit model for each individual night. The bottom panel of each plot shows the residuals between the data and the best fit model.}
\label{fig:RM}
\end{figure}

Before computing this ratio spectrum, the stellar lines of all the spectra, which are shifted in wavelength due to the radial-velocity variation of the star during the transit of a planet, need to be aligned. The radial velocity of WASP-69 is $\pm15~\rm{m~s^{-1}}$, while for HD~189733 is $\pm50~\rm{m~s^{-1}}$. Theses shifts create artificial signals in the spectra if they are not corrected. In order to align the stellar lines to the null stellar radial velocity, we use the HARPS files header information, where the star radial velocity values are given, taking into account possible instrumental, stellar activity and the Rossiter-McLaughlin effects (see Figure~\ref{fig:RM}). Omitting the Rossiter-McLaughlin (RM) effect in the stellar radial velocity when aligning the stellar lines introduces misalignments and spurious results in the final transmission spectrum. 

Some methods obtain the transmission spectrum of the exoplanet atmosphere by dividing the combination of all the telluric-corrected in-transit spectra (master in) by the combination of all the out-of-transit spectra (master out) \citep{2008ApJ...673L..87R}. However, the planet radial velocity varies during the transit, which causes a shift of the planetary absorption lines across the stellar lines profile. The radial velocity of WASP-69b changes from $+7~\rm{km~s^{-1}}$ to $-7~\rm{km~s^{-1}}$ and $\pm15~\rm{km~s^{-1}}$ for HD~189733b, meaning a wavelength shift of $\pm0.14~\rm{\AA}$ and $\pm0.30~\rm{\AA}$, respectively. For $0.01~\rm{\AA}$ steps, like HARPS, this corresponds to more than $10~\rm{pixels}$ shift. Here we correct for this effect by using the method presented in \citet{2015A&A...577A..62W}, in which the planetary signal is shifted to the null radial velocity at the planet rest frame (i.e. the radial velocity in the middle of the transit). Each in-transit spectrum ($F_{\rm{in}}$) is divided by the master-out spectrum, $M_{\rm{out}} = \sum_{out} F_{\rm{out}}(\lambda)$, and the result is shifted to the planet rest frame. Finally, all of these individual transmission spectra are combined, 

\begin{ceqn}
\begin{align}
\Re = \sum_{in} \dfrac{F_{\rm{in}}(\lambda)}{M_{\rm{out}}}\Bigr|_{\substack{\rm{Planet~RV~Shift}}} - 1
\end{align}
\end{ceqn}

This process is applied to all nights, obtaining a transmission spectrum for each night, which are then combined and normalized using a linear fit to the continuum outside the regions of interest.

\begin{figure}
\centering
\includegraphics[width=0.49\textwidth]{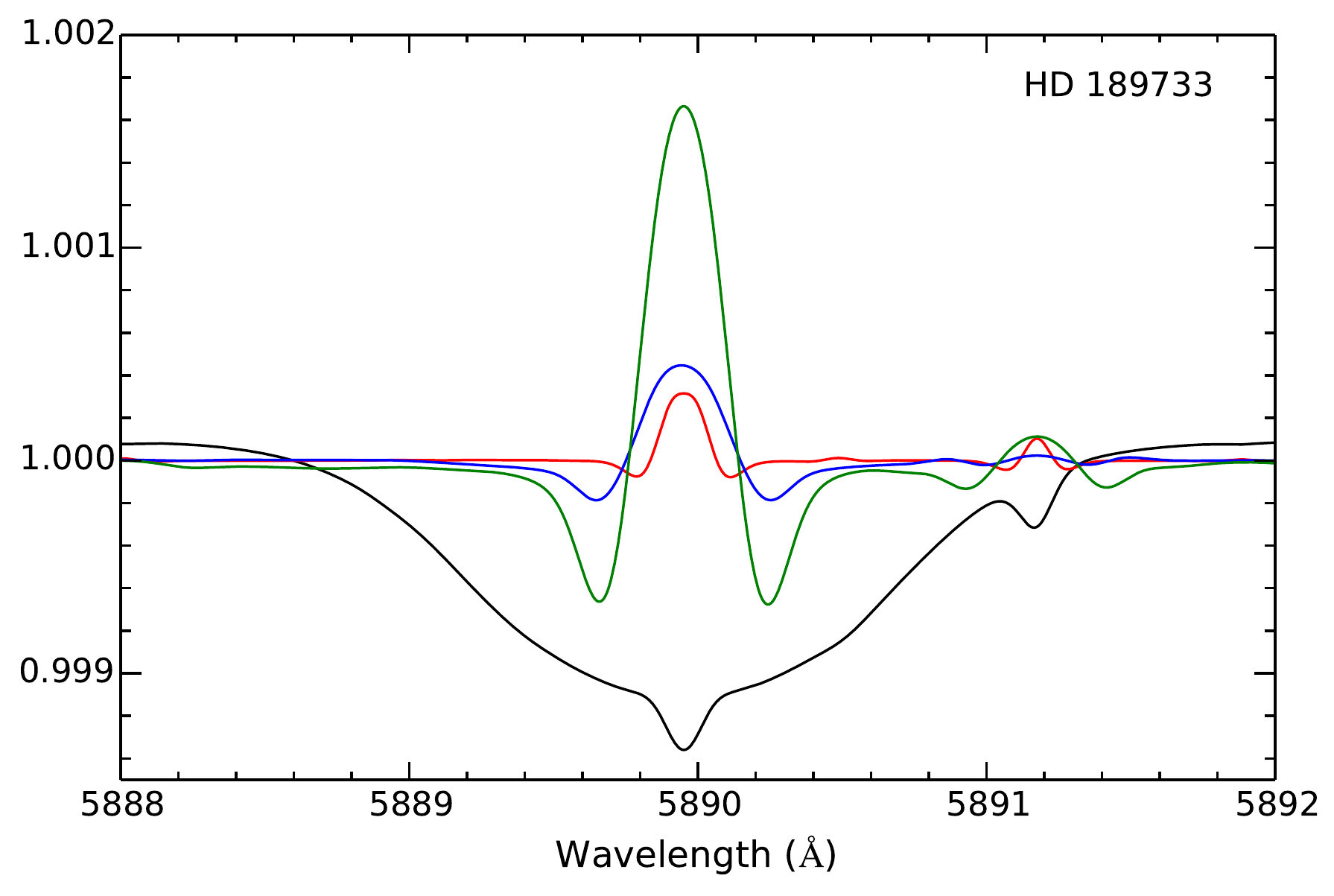}
\caption{Modeled CLV and RM effects of the HD~189733 D2 Na I line. The black line shows the modeled CLV effect and the red line shows the RM effect in the stellar rest frame. The RM effects in the planetary rest frame, with and without correcting the RM induced radial velocity, are shown in blue and green, respectively. For the D1 line, the effects are slightly stronger.}
\label{fig:mod_RM_CLV}
\end{figure}

The CLV effect and RM effect both have impacts on the obtained transmission spectrum. As shown in \citep{2017A&A...603A..73Y}, the center-to-limb variation of the stellar lines profile along the stellar disk is an important effect for the transmission spectroscopy which needs to be corrected for a better characterization of the planetary Na I absorption. Omitting the Rossiter-McLaughlin (RM) effect in the stellar radial velocity (RV) when aligning the stellar lines introduces misalignments and spurious results in the final transmission spectrum (\citealt{2015ApJ...814L..24L}, \citealt{2016MNRAS.462.1012B}). Note that during the transit the planet occults different parts of the stellar disk, which have different velocities. Thus, in addition to the lines shift, the RM also affects the stellar lines shape, which can produce false-positive features in the transmission spectrum. The RM effect can be averaged out if the spectra are retrieved uniformly during a full transit and shifted to the stellar rest frame. However, if the spectra are shifted into the planetary rest frame, the RM effect still has a residual feature. This residual feature is large when the RM induced RV is not corrected but is much smaller when the RV has been corrected (see Figure~\ref{fig:mod_RM_CLV}).

We model the CLV and RM effects together for the case in which the stellar RV (including the RM induced RV) is corrected and the spectra are shifted to the planetary rest frame. We use a similar model method as presented in \citep{2017A&A...603A..73Y}. The CLV and RM effects are then corrected dividing the obtained transmission spectrum by the modeled result.

\section{Results and discussion}

We are interested in the detection of the transitions produced by the Na I doublet at $\lambda5889.951~\rm{\AA}$ (D2) and $\lambda5895.924~\rm{\AA}$ (D1). As the systemic velocity of the planetary system is $-9.62826~\mathrm{km~s^{-1}}$ in the case of WASP-69 system and $-2.2765~\mathrm{km~s^{-1}}$ for HD~189733, the expected wavelength positions of the Na I D lines in the solar system barycentric reference frame are $\lambda5889.762~\rm{\AA}$ (D2) and $\lambda5895.735~\rm{\AA}$ (D1), and $\lambda5889.906~\rm{\AA}$ (D2) and $\lambda5895.879~\rm{\AA}$ (D1), respectively.

In order to compute the relative absorption depths we calculate the weighted mean of the flux in a central pass-band (C), centered on each Na I D lines, and compare it to bins of similar bandwidths taken in the transmission spectrum continuum, one in the blue (B) and one in the red region (R), as presented in \citet{2017A&A...602A..36W},

\begin{ceqn}
\begin{align}
\delta (\Delta\lambda) = \dfrac{\sum_Cw_i\Re(\lambda_i)}{\sum_Cw_i} -  \dfrac{1}{2} \left(\dfrac{\sum_Bw_i\Re(\lambda_i)}{\sum_Bw_i}+\dfrac{\sum_Rw_i\Re(\lambda_i)}{\sum_Rw_i}\right)
\end{align}
\end{ceqn}

where the weights, $w_i$, are the inverse of the squared uncertanties on $\Re$, and $w_i = 1/\sigma_i^2$, which are calculated by propagating the photon noise and the readout noise from the observed spectra.

We choose the same bandwidths $(\Delta\lambda)$ of the central pass-bands used in \citet{2015A&A...577A..62W} for easier comparison: $0.188~\mathrm{\AA}$, $0.375~\mathrm{\AA}$, $0.75~\mathrm{\AA}$, $1.5~\mathrm{\AA}$, $3~\mathrm{\AA}$ and a larger pass-band of $12~\mathrm{\AA}$ adjusted on the center of the doublet to perform the comparison between our results and those of \citet{2008ApJ...673L..87R}, \citet{jensen2011} and \citet{huitson2012}. In this last case, as reference we use slightly different wavelength intervals: $5874.89-5886.89~\mathrm{\AA}$ corresponding to the blue and $5898.89-5907.89~\mathrm{\AA}$ to the red regions. As noted by \citep{2017A&A...602A..36W}, we also find that the choice of the reference bands has no impact on the results.

\begin{figure*}[]
\centering
\includegraphics[width=\textwidth]{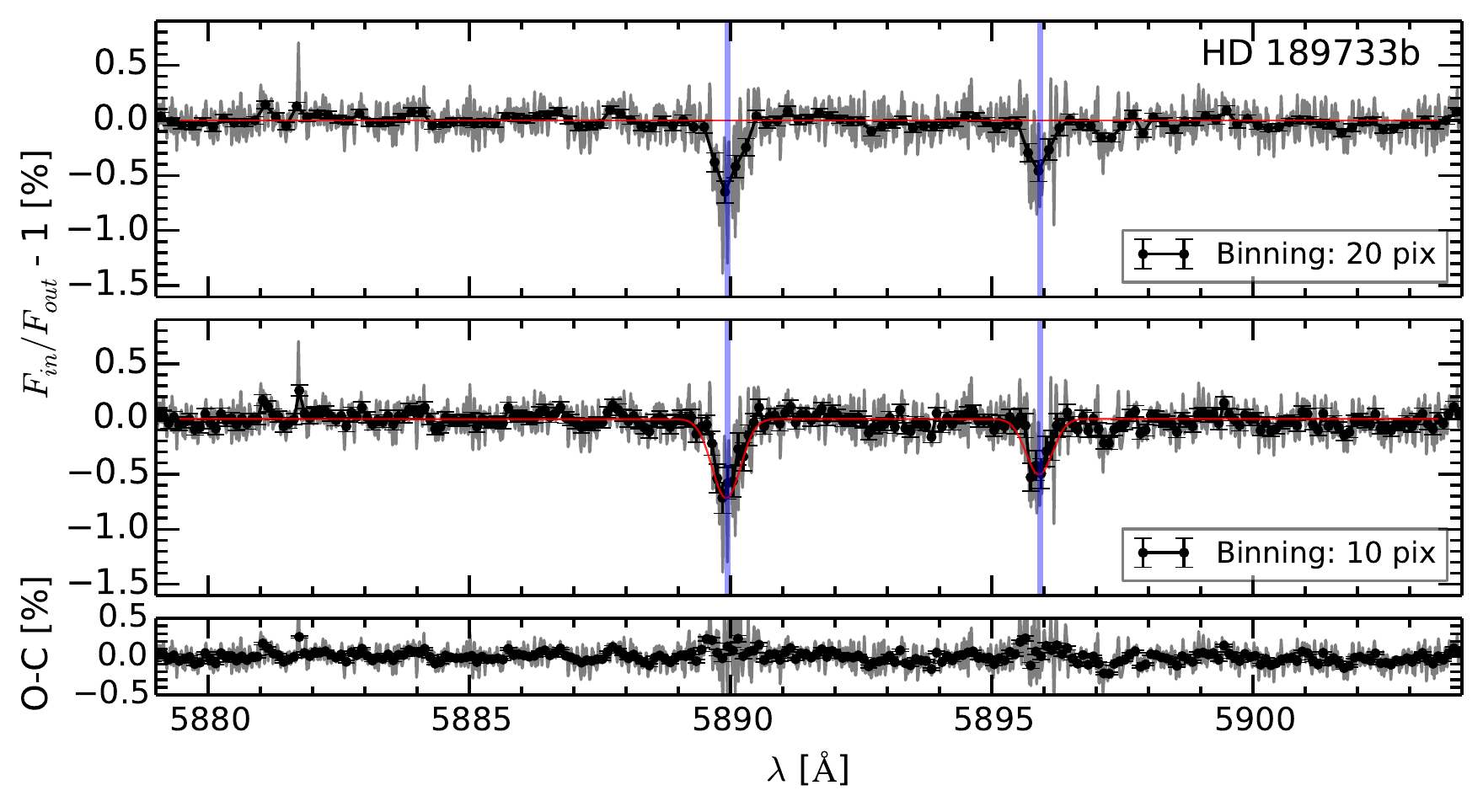}
\includegraphics[width=\textwidth]{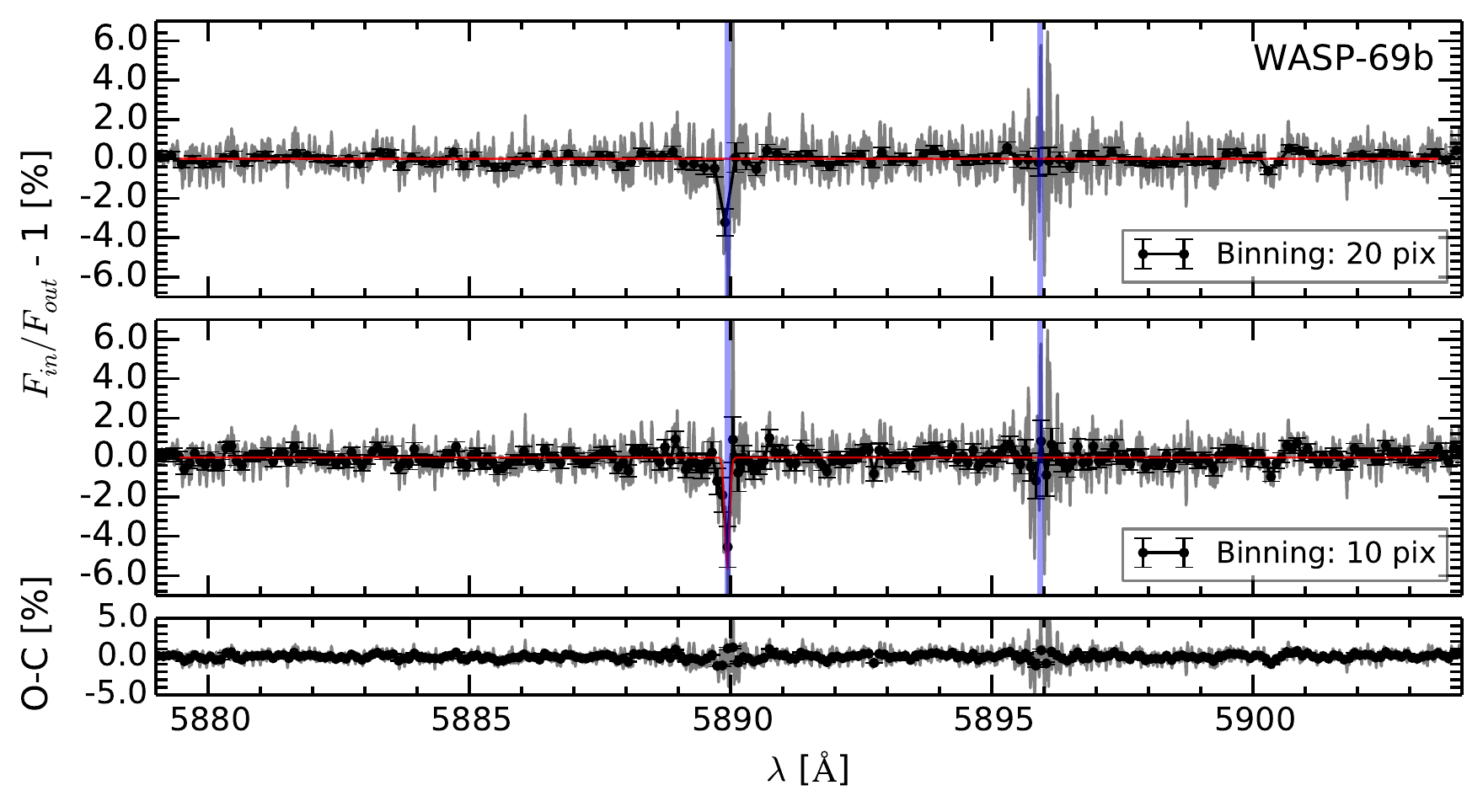}
\caption{Transmission spectrum of HD~189733b (top panel) and WASP-69b (bottom panel) atmospheres in the region of Na I D doublet. The atmospheric transmission spectrum is presented in light grey. With black dots, we show the binned transmission spectrum, 20 pixels (top) and 10 pixels (middle). The gaussian fit to each Na I D lines is shown in red with its residuals in the bottom. The expected wavelength position of the Na I doublet lines, in the planetary reference frame, are indicated with blue vertical lines and the red horizontal line is used as reference around zero. A net blueshift of ${\sim}3~\rm{km~s^{-1}}$ (${\sim}0.04\rm{\AA}$) is observed in the Na I lines of HD~189733b, possibly due to a global error on the planetary radial velocity shift as a result of the HARPS precision. No shift is observed for WASP-69b Na I lines with respect the reference wavelength frame. The uncertainties of the relative flux are assumed to come from photon and readout noise and propagated from the original spectra.}
\label{fig:results}
\end{figure*}

\subsection{Transmission spectrum analysis}

The signal-to-noise ratio (SNR) per extracted pixel in the continuum near $590~\mathrm{nm}$ for HD~189733, retrieved with HARPS-South, ranges from $55-240$ for all nights, while for WASP-69 spectra, with HARPS-North, the SNR is lower: $30$-$60$. Combining the spectra to obtain the master in and master out spectra, the SNR increases to $310-640$  for HD~189733 and $100-120$ for WASP-69. When co-adding the nights, the total SNR for each master is about $800$ for HD~189733 and $150$ for WASP-69. Thus, a SNR of $1500$ and $205$ is reached in the final transmission spectrum for HD~189733b and WASP-69b, respectively. 

As observed in Figure~\ref{fig:results}, for HD~189733b, both Na I D lines can be distinguished from the continuum noise. However, for WASP-69b, even if the D2 line clearly peaks out the continuum, the D1 line  easily disappears when the resulting transmission spectrum data is binned with a large number of pixels. With a gaussian fit to each Na I lines, we measure line contrasts of $0.72\pm0.05~\%$ (D2) and $0.51\pm0.05~\%$ (D1) and FWHMs of $0.64\pm0.04~\mathrm{\AA}$ (D2) and $0.60\pm0.06~\mathrm{\AA}$ (D1) for HD~189733b. Similar results of line contrasts are presented in \citet{2015A&A...577A..62W}, but larger values of the FWHMs are obtained here. For WASP-69b, taking into account the low SNR reached, a gaussian fit in the D1 Na I line is not possible. For the D2 line, even if the gaussian fit is better and the line contrast can be measured ($5.8\pm0.3~\%$), more transits are needed to estimate the FWHM.

For HD~189733b, a small net blueshift of ${\sim}-0.04~\mathrm{\AA}$ is measured with respect to the expected wavelength position of the Na I lines. Taking into account that one pixel in the HARPS detector represents ${\sim}0.8~\mathrm{km~s^{-1}}$, the measured shift corresponds to a ${\sim}-2~\rm{km~s^{-1}}$. This value is consistent with the wind speed detected by \citet{Brogi2016} ($-1.7^{+1.1}_{-1.2}$) using data from CRIRES (IR) and \citet{2015ApJ...814L..24L} ($-1.9^{+0.7}_{-0.6}$) using the same optical data analyzed here, suggesting almost no vertical wind shear. This shift is by far smaller than the ${\sim}0.75~\mathrm{\AA}$ presented in \citet{2008ApJ...673L..87R} and smaller than the $0.16~\mathrm{\AA}$ measured in \citet{2015A&A...577A..62W}. In this last paper speculate that a global error on the planetary radial-velocity shift could introduce a $0.06~\mathrm{\AA}$ shift, considering the HARPS precision. For WASP-69b no blueshift is measured.

We measure relative absorption depth of the Na I lines in the transmission spectra for different passbands. For WASP-69b only the D2 Na I line is used while for HD~189733b the absorption depth of both lines is averaged. The values obtained for HD~189733b are $0.269\pm0.032~\%$ ($8.5\sigma$) and $0.030\pm0.014~\%$ ($2.1\sigma$) for $0.75~\mathrm{\AA}$ and $12~\mathrm{\AA}$ bandwidths, respectively (see first row Table~\ref{tab:adlines_values}). Comparing these results with those in \citet{2015A&A...577A..62W} we obtain slightly smaller values possibly due to the differences between the data analysis methods. For the $12~\mathrm{\AA}$ pass-band our results are smaller than the ones presented in this last paper ($0.056\pm0.007~\%$), in \citet{jensen2011} ($0.053\pm0.017~\%$) and \citet{2008ApJ...673L..87R} ($0.067\pm0.020~\%$) from ground. Also in the detection from space by \citet{huitson2012} ($0.051\pm0.006~\%$) the results are larger than ours. However, taking into account the uncertainties of the values, the differences are not significant. 

On the other hand, for WASP-69b we measure $3.819\pm0.764$ ($5\sigma$), $2.058\pm0.491$ ($4.2\sigma$), $0.703\pm0.284$ ($2.5\sigma$), $0.529\pm0.142$ ($3.7\sigma$) and $0.114\pm0.080$ ($1.5\sigma$) for $0.188~\mathrm{\AA}$, $0.375~\mathrm{\AA}$, $0.75~\mathrm{\AA}$, $1.5~\mathrm{\AA}$, $3~\mathrm{\AA}$ pass-bands, respectively (see second row of Table~\ref{tab:adlines_values}). Note that since the $12~\mathrm{\AA}$ pass-band includes both Na I lines and only the D2 line is used for WASP-69b, the absorption depth for this bandwidth is not calculated.

\begin{figure}[h]
\centering
\includegraphics[width=0.49\textwidth]{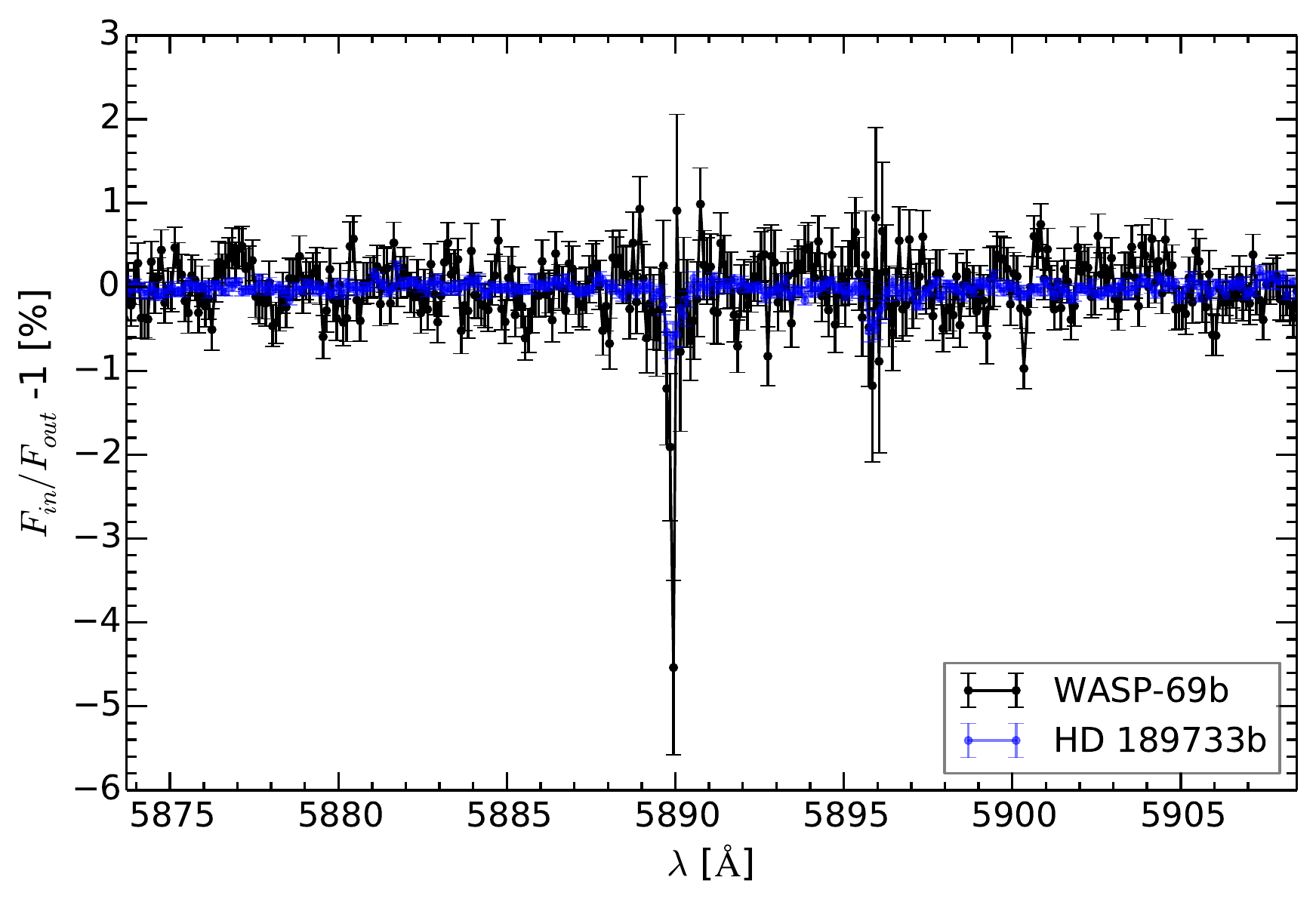}
\caption{Comparison between the transmission spectrum of WASP-69b (black) and HD~189733b (blue) in the region of the Na I doublet. Both transmission spectra are shown binned by 10 pixels and the wavelength shifted to the planet reference frame.}
\label{fig:comp_TS}
\end{figure}

It is important to emphasize that WASP-69 is two magnitudes fainter than HD~189733. In order to extract the transmission spectrum of WASP-69b, two transits of this planet with approximately 15 spectra each were combined, while for HD~189733b three transits with 40 spectra, on average, were available. The fact that we can see Na absorption in WASP-69b's transmission spectrum means that the signal must be stronger than that from HD~189733b's. Indeed, the scale height of WASP-69b is ${\sim}650~\mathrm{km}$, three times larger than HD~189733b's (${\sim}200~\mathrm{km}$), see Figure~\ref{fig:comp_TS}. Nevertheless, more WASP-69b transit spectra will be needed to reach the enough SNR to verify the D1 line detection and characterize the Na I lines profile. At present, there is no clear explanation for the difference in the intensity of both Na I lines. These lines are formed very high in the atmosphere and the theoretical lines profiles of the two lines are different, presenting different oscillation strengths ($f=0.64$ for the D2 and $f=0.32$ for the D1) \citep{huitson2012}. As suggested in \citet{2000Slanger}, the variability of the D2/D1 ratio could be due to different pathways, creating the two excited Na I levels with different temperature sensitivities. This would result in a different Na I lines intensity depending on the local conditions in the exoplanet atmosphere. On Earth, for example, the telluric Na I absorption of D2 line is stronger than D1 line \citep{thesisSimon}.

\begin{figure}[h]
\centering
\includegraphics[width=0.49\textwidth]{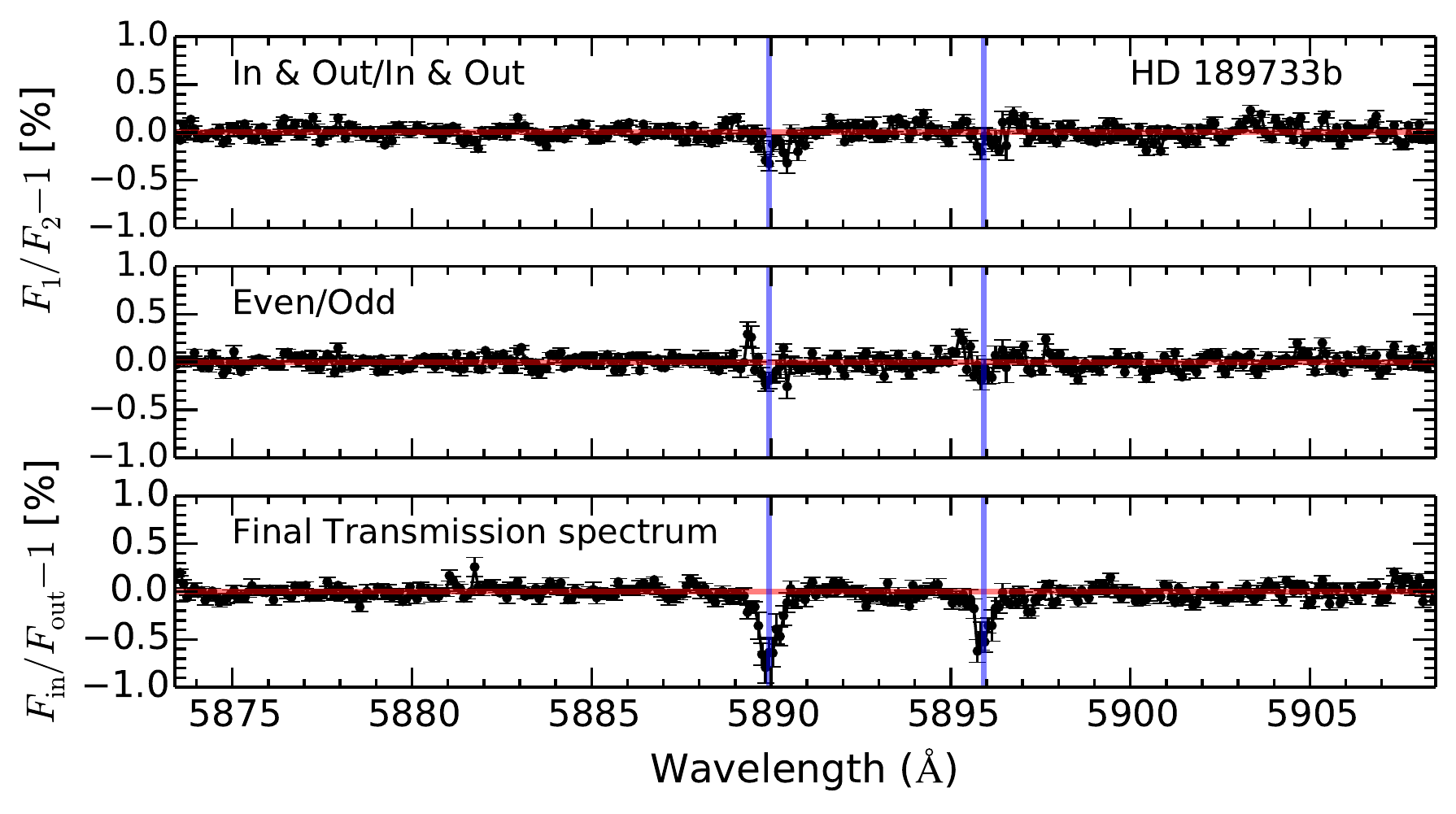}
\includegraphics[width=0.49\textwidth]{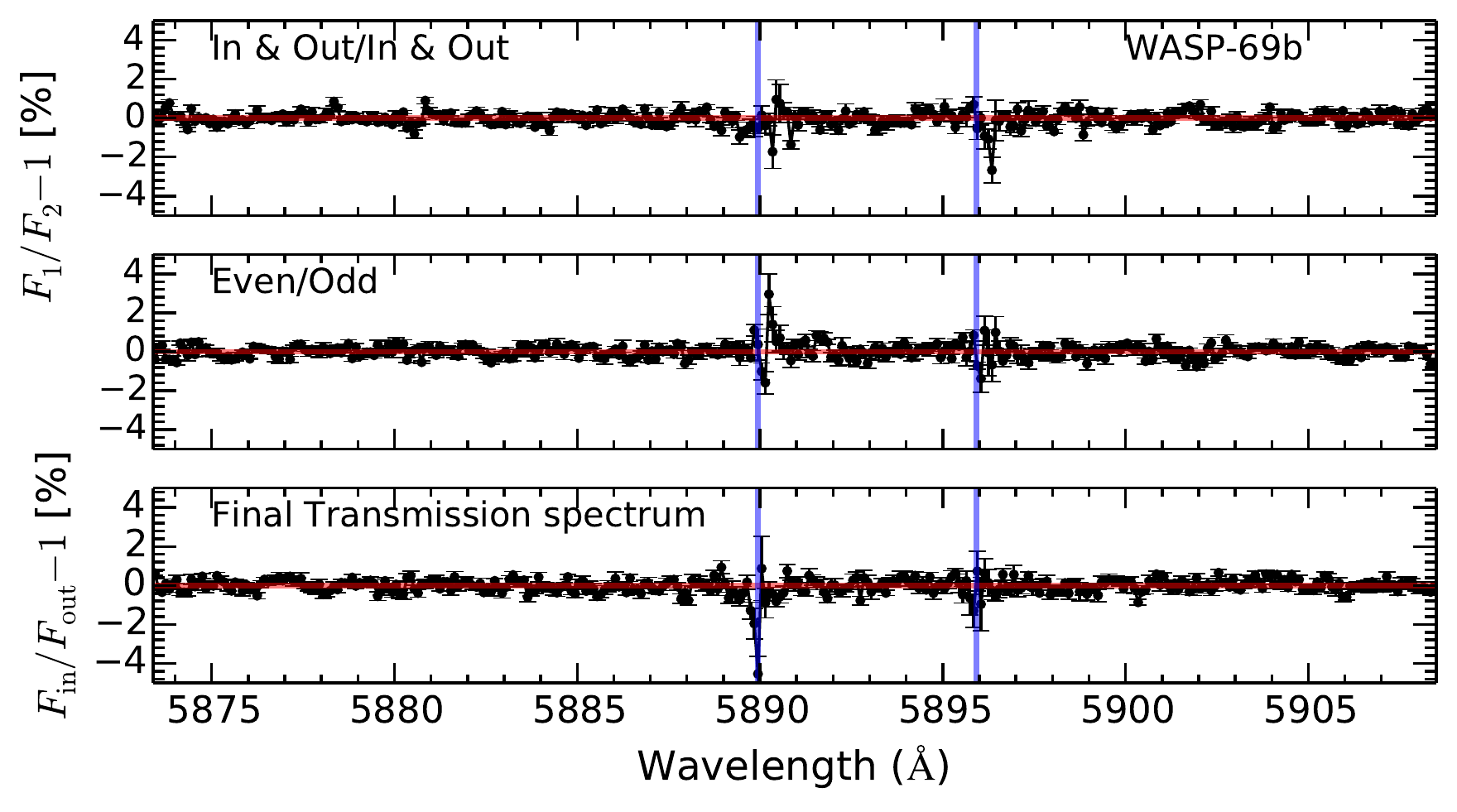}
\caption{Mock transmission spectra obtained with different combination of in- ($F_{1}$) and out-of-transit ($F_{2}$) files of HD~189733 (top panel) and WASP-69 (bottom panel). The results are shown binned every 10 pixels. The horizontal red line is used as reference around zero and the blue vertical lines are in the expected wavelength position of the Na I lines in the planet reference frame. First row: final transmission spectrum obtained using a synthetic in-transit sample formed by the data before transit and the second half of in-transit data (ordered according to time) and the rest as a synthetic out-of-transit sample. Second row: transmission spectrum obtained considering the even files as the in-transit sample and the odd ones as the out-of-transit sample. Third row: Final transmission spectrum from our data reduction.}
\label{fig:checks}
\end{figure}

In order to be sure that the results obtained are not caused by stellar residuals, some checks were performed to prove that the resulting transmission spectrum can only be obtained with the correct selection of the in- and out-of-transit samples. In particular, we computed the transmission spectrum by considering different combination of in- ($F_{1}$) and out-of-transit ($F_{2}$) files and following the same reduction steps. For the first check, we used the before-transit files and the second half of the in-transit files (ordered according to time) as a synthetic in-transit sample, and the rest of the files (first half of in-transit and after-transit files) as a synthetic out-of-transit sample (see first row of Figure~\ref{fig:checks}). The second test was performed by considering the even files forming the in-transit sample and the odd ones (also organized by time) the out-of-transit sample (see second row of Figure~\ref{fig:checks}).

\renewcommand{\thefootnote}{\fnsymbol{footnote}}
 \begin{table*}[ht]
\centering
\caption{Summary of the measured relative absorption depth in [$\%$] in the transmission spectra of HD~189733b and WASP-69b. Both D2 and D1 Na I lines are averaged for HD~189733b, and only the D2 line is used for WASP-69b.}
\begin{tabular}{lcccccc}
\hline\hline
& & & Bandwidth  & & &\\
\cline{2-7}
&$0.188~\mathrm{\AA}$ & $0.375~\mathrm{\AA}$& $0.75~\mathrm{\AA}$&$1.5~\mathrm{\AA}$ &$3.0~\mathrm{\AA}$ &$12~\mathrm{\AA}$\\
\hline
HD~189733b & $0.506\pm0.074$ & $0.497\pm0.050$ &$0.269\pm0.032$ & $0.104\pm0.017$ & $0.059\pm0.013$ & $0.030\pm0.014$\\
\hline
WASP-69b & $3.819\pm0.764$& $2.058\pm0.491$ & $0.703\pm0.284$ & $0.529\pm0.142$ & $0.114\pm0.080$ & $-$\footnote[1]{}\\
\hline
\end{tabular}\\
\begin{tablenotes}
\item Notes.$^($\footnotemark[1]$^)$ The bandwidth of $12~\mathrm{\AA}$ includes both Na I lines and only the D2 Na I line is used to compute the absorption depths of WASP-69b. 
\end{tablenotes}
\label{tab:adlines_values}
\end{table*}
\renewcommand{\thefootnote}{\arabic{footnote}}

In both checks, the weight of real in-transit and out-of-transit files considered in $F_{1}$ and $F_{2}$ samples is the same, and we expect a flat transmission spectrum, despite possible stellar residuals caused by stellar activity, the Rossiter-McLaughlin effect (which is not completely averaged now), the center-to-limb variation, among other possible sources of change in the stellar lines. For HD~189733b the resulting synthetic transmission spectra are mainly flat, only some strong residuals in the Na I lines position are observed, but with clear difference with the lines depth of the real transmission spectrum. For WASP-69b stronger residuals near the Na I position are observed. However, while the planetary radial-velocity correction has been applied to each single $F_{1}/F_{2}$, none of these residuals are in the position of the planetary Na I absorption, and are very likely of stellar in origin. These stellar residuals illustrate the importance of correcting for the stellar line shifts, which can easily become false-positives in the final transmission spectrum. 

As discussed in the reduction process, missing some of the reduction steps could introduce residuals in the final transmission spectrum. Some of these steps are the telluric Na correction, the barycentric Earth radial velocity consideration in the telluric lines subtraction, the RM induced radial velocity in the stellar lines alignment, and the correction of the line's shape caused by CLV and RM effects. These residuals have been quantified by measuring the absorption depth values of the transmission spectrum of WASP-69b and HD~189733b by omitting these steps in the reduction process. The absorption depth values obtained for a representative $1.5~\mathrm{\AA}$ passband are shown in Table~\ref{tab:AD_meth}. The most important differences are observed omitting the telluric Na correction, for WASP-69b, and not considering the RM induced radial velocity for HD~189733b. 

\renewcommand{\thefootnote}{\fnsymbol{footnote}}
 \begin{table*}[ht]
\centering
\caption{Measured relative absorption depth in [$\%$] in the transmission spectra of HD~189733b and WASP-69b omitting steps in the reduction process, for a $1.5~\mathrm{\AA}$ bandwidth. Both D2 and D1 Na I lines are averaged for HD~189733b, and only the D2 line is used for WASP-69b.}
\begin{tabular}{lccccc}
\hline\hline
& & & Omitted correction & & \\
\cline{3-6}
& Result\footnote[6]{} &Telluric Na\footnote[1]{}& BERV in telluric lines\footnote[2]{} & RM induced RV\footnote[3]{} & CLV+RM effects\footnote[4]{} \\
\hline
HD~189733b & $0.104\pm0.017$&$-$\footnote[5]{} & $0.185\pm0.018$ &$0.196\pm0.017$ & $0.188\pm0.017$ \\
\hline
WASP-69b & $0.529\pm0.142$ & $0.628\pm0.139$& $0.467\pm0.142$ & $0.492\pm0.143$ & $0.584\pm0.142$ \\
\hline
\end{tabular}\\
\begin{tablenotes}
\item Notes.$^($\footnotemark[6]$^)$Measured absorption depth in the final transmission spectra following all the presented steps. $^($\footnotemark[1]$^)$ No telluric Na subtraction.  $^($\footnotemark[2]$^)$The barycentric Earth radial velocity consideration in the telluric lines subtraction. $^($\footnotemark[3]$^)$RM induced radial velocity consideration in the stellar lines alignment. $^($\footnotemark[4]$^)$Correction of the line's shape caused by CLV and RM effects. $^($\footnotemark[5]$^)$ No telluric Na is observed in HD~189733 spectra.
\end{tablenotes}
\label{tab:AD_meth}
\end{table*}
\renewcommand{\thefootnote}{\arabic{footnote}}

\subsection{(Spectro-)Photometric light curve of Na I}

\begin{figure}[h]
\centering
\includegraphics[width=0.49\textwidth]{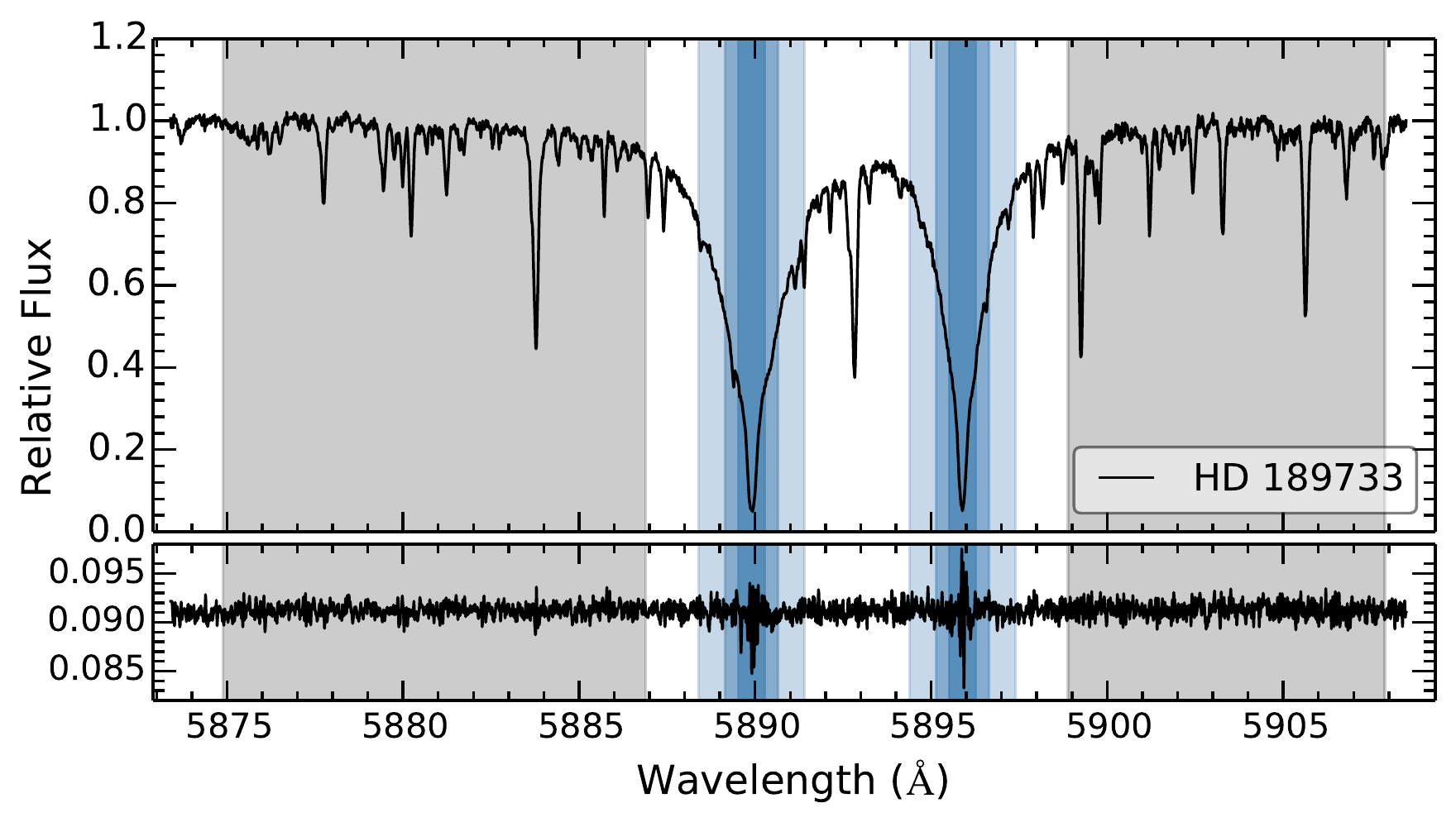}
\includegraphics[width=0.49\textwidth]{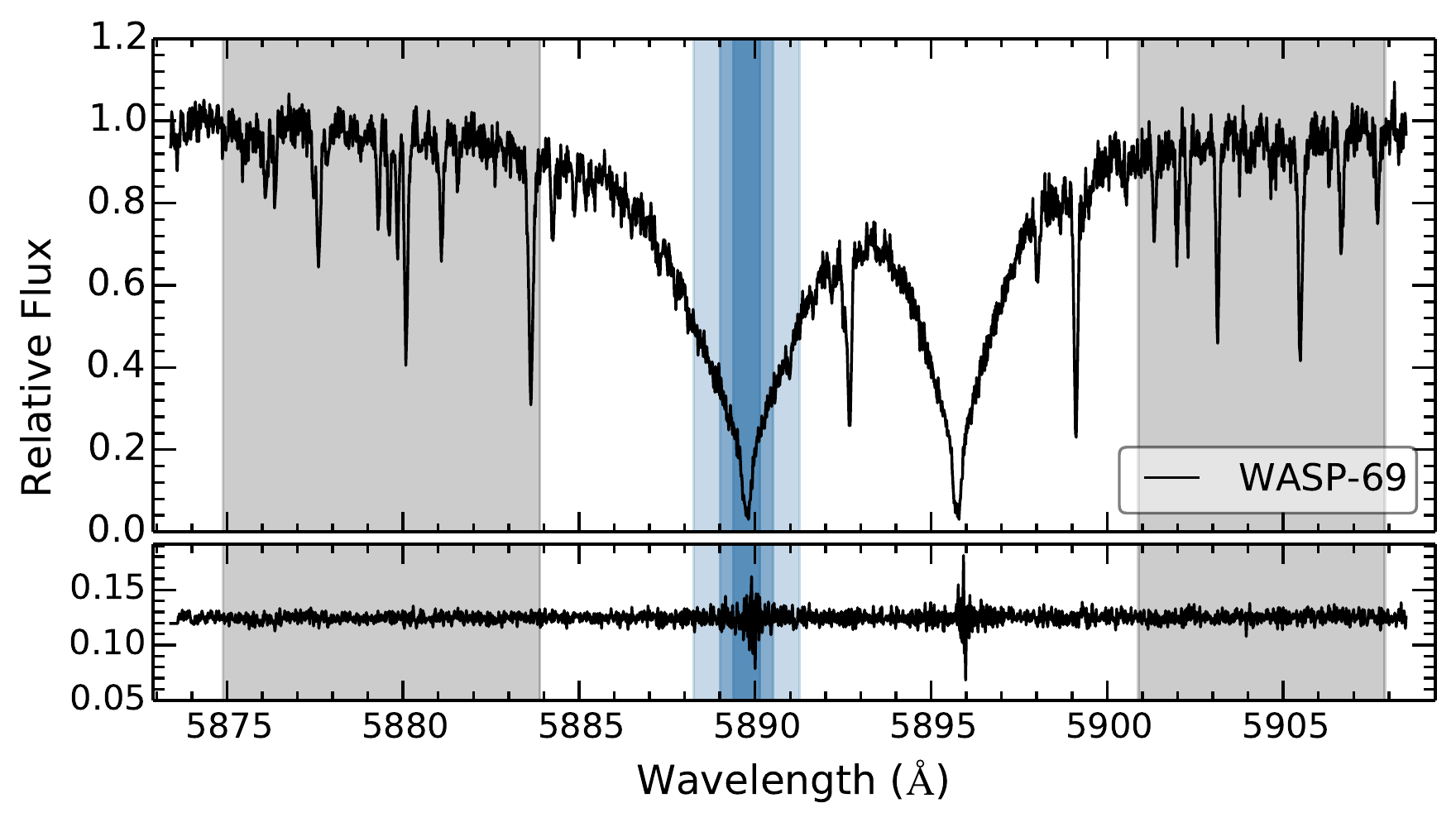}
\caption{Passbands used to perform the transmission light curves of HD~189733b (upper panel) and WASP-69b (lower panel). One observed spectrum is shown on the top of each panel and the ratio between this single in-transit spectrum and the master out is shown on the bottom.  The blue regions mark the three different bandwidths ($0.75~\rm{\AA}$, $1.5~\rm{\AA}$ and $3.0~\rm{\AA}$) centered at the line cores of D1 and D2, and the blue and red reference passbands used to measure the relative flux and calculate the transmission light curves are shown in gray. Note that for WASP-69 only the D2 line is used.}
\label{fig:passbands}
\end{figure}

The transit light curve of the Na I line is calculated using three different bandwidths: $0.75~\rm{\AA}$, $1.5~\rm{\AA}$ and $3.0~\rm{\AA}$, which are centered at the line cores of each D1 and D2 lines, after dividing each spectrum (in- and out-of-transit, $F_{\mathrm{in}}$ and $F_{\mathrm{out}}$) by the high resolution master out ($M_{\rm{out}}$). As reference passbands, for HD~189733b, we use similar ranges as in \citet{2015A&A...577A..62W} and \citet{2017A&A...603A..73Y}, $5874.89~\mathrm{\AA}{\sim}5886.89~\mathrm{\AA}$ for the blue region and $5898.89~\rm{\AA}{\sim}5907.89~\rm{\AA}$ for the red region. For WASP-69b, as the Na I lines are wider (see Figure \ref{fig:example_spec}), the reference passbands are defined as $5874.89~\rm{\AA}{\sim}5883.89~\rm{\AA}$ for the blue part and $5900.89~\rm{\AA}{\sim}5907.89~\rm{\AA}$ for the red part (see Figure~\ref{fig:passbands}).

\begin{figure*}[]
\centering
\includegraphics[width=\textwidth]{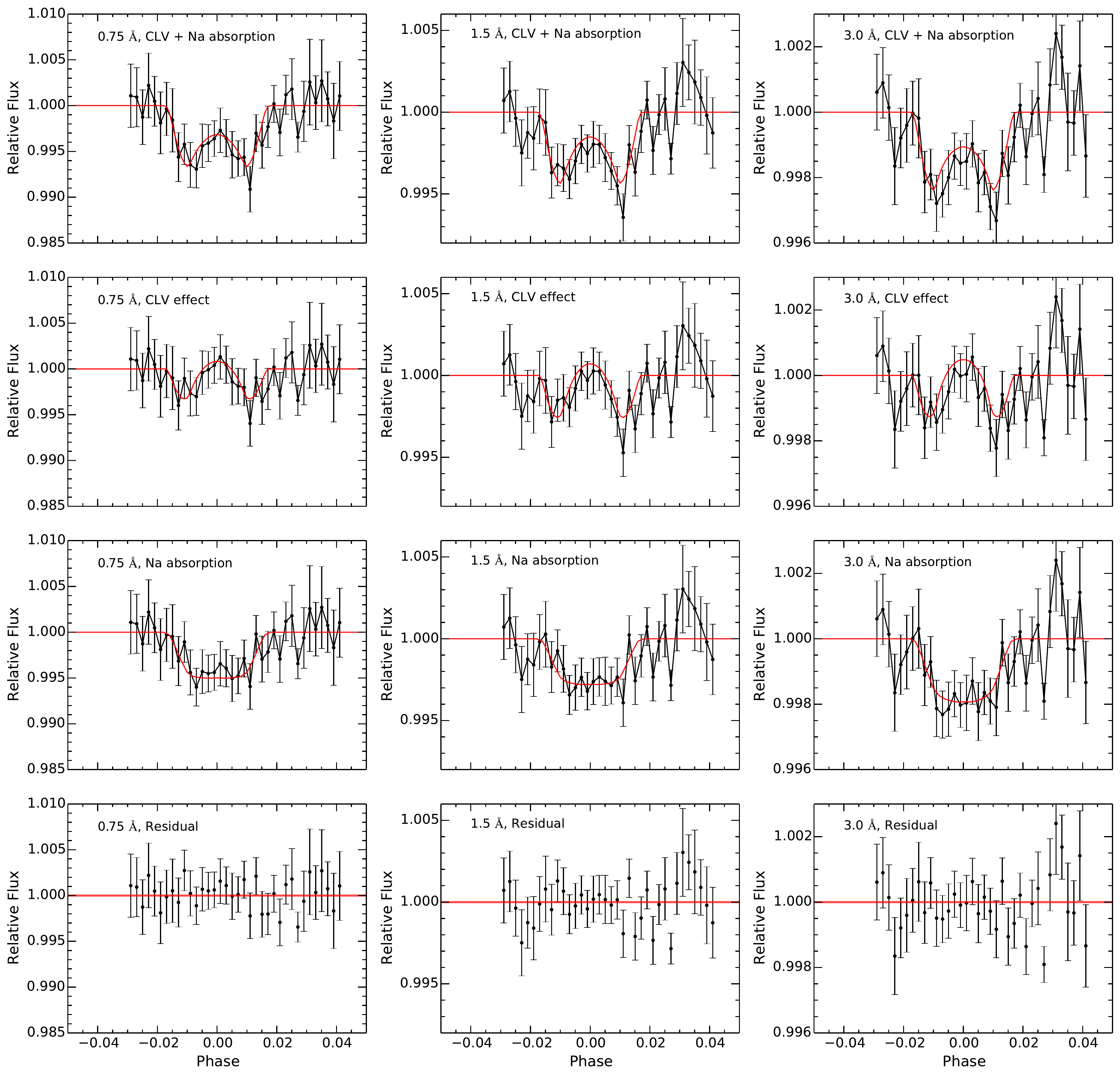}
\caption{Transmission light curves from HARPS observations of HD~189733b, for three different bandwidths: $0.75~\rm{\AA}$ (left column), $1.5~\rm{\AA}$ (middle column) and $3.0~\rm{\AA}$ (right column). Only Night 1 and 3 are used here. First row: Final observed transmission light curve of the Na I D lines from our data reduction. The relative flux of D1 and D2 lines is averaged and binned with a $0.002$ phase step (black line). The red line is the modeled transmission light curve, which is the combination of the best-fit Na I absorption and the CLV effect \citep{2017A&A...603A..73Y}. This model is calculated theoretically and there is no fitting to the actual data. Second row: The CLV effect. The black line is the result of removing the Na I absorption model from the observed transmission light curve. The modeled CLV effect is shown in red. Third row: The Na I absorption light curve obtained dividing the observed transmission light curve by the modeled CLV effect (black line). The red line is the best-fit Na I absorption model. Forth row: Residuals between the observed transmission light curve and the model. The errorbars represent the photon noise within the passband.}
\label{fig:light_curves_HD189}
\end{figure*}

\begin{figure*}[]
\centering
\includegraphics[width=\textwidth]{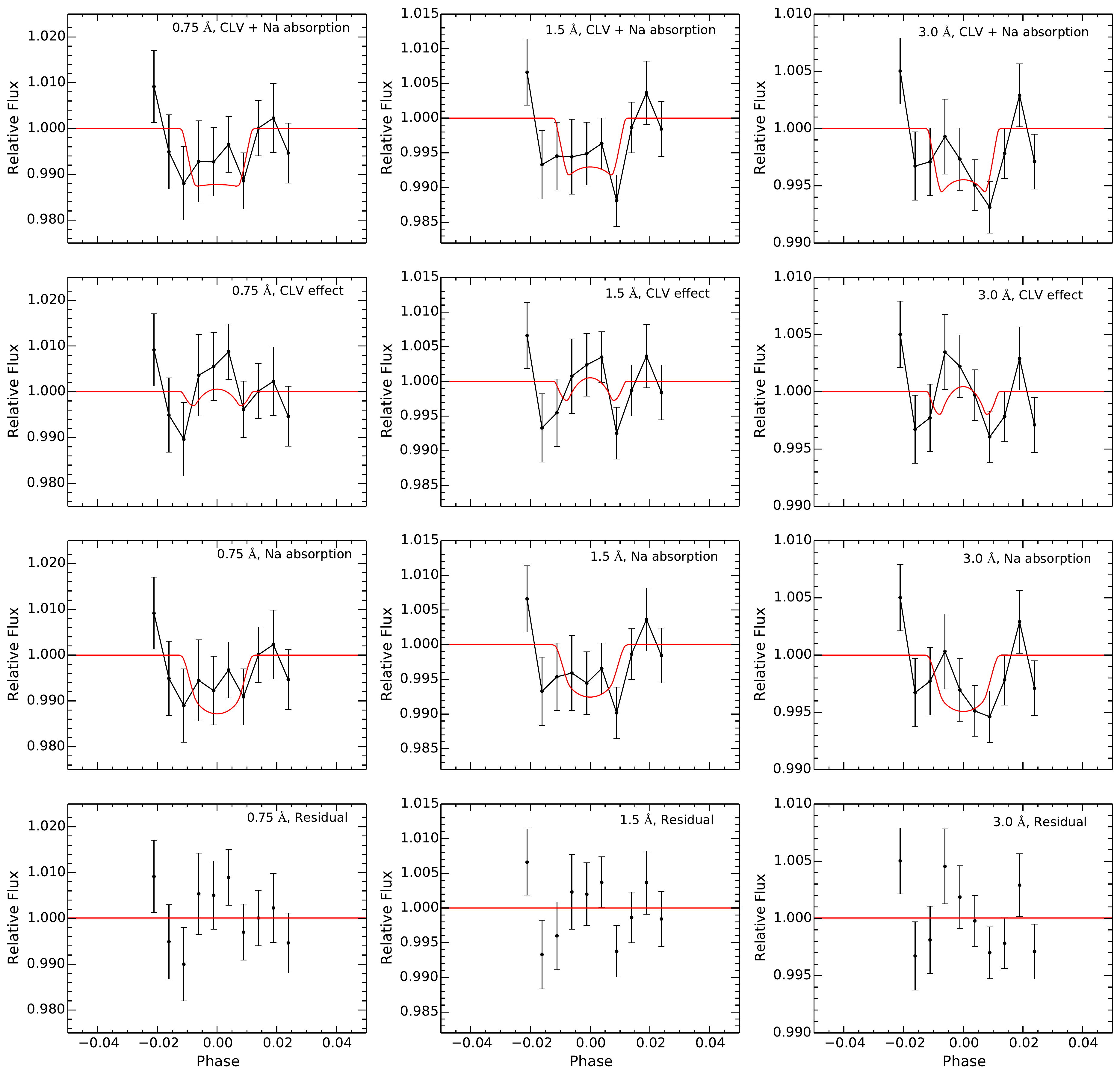}
\caption{Same as Figure~\ref{fig:light_curves_HD189} but for the HARPS-N observations of WASP-69b. In this case, only the relative flux of D2 line is used and binned with a $0.005$ phase step (black line).
}
\label{fig:light_curves_W69}
\end{figure*}

In order to obtain the transit light curve of Na I, we use the method presented in \citet{2017A&A...603A..73Y}, where the relative flux of the Na I lines ($F_{\rm{line}}$) is calculated using the ratio between the flux inside the band centered on the line core ($F_{\rm{cen}}$) and the reference bands ($F_{\rm{red}}$ and $F_{\rm{blue}}$) as follow,
\begin{ceqn}
\begin{align}
F_{\rm{line}} = \dfrac{2~F_{\rm{cen}}}{F_{\rm{red}}+F_{\rm{blue}}}
\end{align}
\end{ceqn}
This relative flux is obtained for both D2 and D1 lines in the case of HD~189733b, normalized to unity for all the out-of-transit values and then, the light curves of both lines are averaged. Note that only the first and third nights are used to calculate the relative flux, since for the second night there were no observations before transit. For WASP-69b the relative flux is calculated only for the D2 line. Finally, we binned the data points with a 0.002 phase step for HD~189733b, for an easier comparison with similar results \citep{2017A&A...603A..73Y}. For WASP-69b the observed spectra are more noisy (SNR${\sim}40$) than the retrieved of HD~189733b (SNR${\sim}130$) (see Figure~\ref{fig:example_spec}). This noise is propagated along the calculation of the final transmission light curve values, which present a high dispersion. In order to reduce this dispersion we decided to increase the binned phase step to 0.005 for this planet. The final observed transmission light curves for the three different bandwidths are shown in the first row of Figures~\ref{fig:light_curves_HD189} and ~\ref{fig:light_curves_W69}.

The resulting light curves are the combination of the planetary absorption and the center-to-limb variation effect. Here, we present the observed transmission light curves of the Na I lines which show the characteristic shape of the CLV effect. To correct for this effect we follow the same method as \citet{2017A&A...603A..73Y}, dividing the observed transmission light curve by the modeled CLV effect light curve from the same paper (see second row of Figures~\ref{fig:light_curves_HD189} and \ref{fig:light_curves_W69}). This makes it possible to obtain the observed Na I absorption light curve and calculate the true absorption depth (see third row of Figures~\ref{fig:light_curves_HD189} and \ref{fig:light_curves_W69}). For WASP-69b the modeled Na I absorption is obtained using the PyTransit \citep{Parviainen2015Pytransit} and PyLDTk \citep{Parviainen2015LDTK} Python packages, which use the library of PHOENIX stellar atmospheres \citep{Husser2013}. It is important to note that the modeled CLV light curve is directly derived from the model and the fit procedure is only applied to the Na I absorption model.

The transmitted flux or relative absorption depth of the transmission light curves can be easily calculated by the difference of the weighted mean of the flux in- and out-of-transit:
\begin{ceqn}
\begin{align}
\delta(\Delta\lambda) = \dfrac{\overline{F_{line}(in)}}{\overline{F_{line}(out)}}
\end{align}
\end{ceqn}

 For HD~189733b the observed light curve can be well fitted with the combined model (CLV and Na I absorption), with a chi-square values ($\chi^2$) of $1.3$, in average for the different bandwidths. In \citet{2017A&A...603A..73Y}, a $\chi^2 = 2.6$ with the combined model was reported. In both studies, the $\chi^2$ values obtained by fitting only the Na I absorption model are larger than the ones obtained fitting the combined model ($\chi^2 = 2.0$ in our work). As in \citet{2017A&A...603A..73Y}, we clearly observe the CLV effect in our data, which is not obvious in the light curves presented in \citet{2015A&A...577A..62W}. Similar CLV features are also shown in the light curves presented in \citet{2016MNRAS.462.1012B}. With the CLV effect correction applied, we measure absorption depths of $0.442\pm0.074~\%$, $0.274\pm0.052~\%$ and $0.207\pm0.031~\%$, for the $0.75~\rm{\AA}$, $1.5~\rm{\AA}$ and $3.0~\rm{\AA}$ bandwidths, respectively (see Table~\ref{tab:ad_values}). If the CLV effect is not considered, the absorption depths are overestimated (${\sim}0.508~\%$, ${\sim}0.320~\%$ and ${\sim}0.217~\%$, respectively).

 \begin{table*}[h]
\centering
\caption{Summary of the measured relative absorption depth in [$\%$] in the light curve of  HD~189733b and WASP-69b after the CLV correction. Both D2 and D1 Na I lines are averaged for HD~189733b, while for WASP-69b only the D2 line is used. The results of Nigh 1 and 3 for HD~189733, and both nights for WASP-69b are averaged.}
\begin{tabular}{lccc}
\hline\hline
& & Bandwidth  &\\
\cline{2-4}
& $0.75~\mathrm{\AA}$&$1.5~\mathrm{\AA}$ &$3.0~\mathrm{\AA}$ \\
\hline
HD~189733b & $0.442\pm0.074~\%$ & $0.274\pm0.052~\%$ & $0.207\pm0.031~\%$\\
\hline
WASP-69b & $0.722\pm0.411~\%$& $0.470\pm0.235~\%$ & $0.286\pm0.136~\%$\\
\hline
\end{tabular}\\
\label{tab:ad_values}
\end{table*}

Comparing the results with those in \citet{2015A&A...577A..62W}, we find that our absorption depths (without the CLV correction) are larger than theirs. In the case of the $0.75~\mathrm{\AA}$ passband, for example, the averaged absorption depth is $0.40~\%$, while the results agree more closely with the ones presented in \citet{2016MNRAS.462.1012B} ($0.45~\%$) and \citet{2017A&A...603A..73Y} ($0.512~\%$). Comparing the absorption depths measured after the CLV correction, in \citet{2017A&A...603A..73Y} measure $0.414\pm0.032~\%$ for the $0.75~\rm{\AA}$ bandwidth, slightly different than our result ($0.442\pm0.074~\%$). The differences probably arise from the different reduction methods used, and fall within the uncertainties.

As previously discussed, the number of spectra retrieved during the observations of WASP-69b is small compared to the large number of spectra used to perform the HD~189733b transmission light curves. Thus, the resulting transmission light curves of WASP-69b are noisy and the combined model does not fit the observed data as well as for HD~189733b ($\chi^2  = 4.2$). Even if the CLV is not obvious in the results because of the noise, this effect is predicted in our model and used to correct the observed transmission light curve. In this case, the fit is also better if both contributions (Na I absorption plus CLV) are considered. We measure absorption depths of $0.722\pm0.411~\%$, $0.470\pm0.235~\%$ and $0.286\pm0.136~\%$, for the $0.75~\rm{\AA}$, $1.5~\rm{\AA}$ and $3.0~\rm{\AA}$ bandwidths respectively. Note that also larger values are obtained without considering the CLV effect.  

In all cases, the measurement uncertainties are calculated with the error propagation of the photon noise level and the readout noise from the original data.  

\section{The Rossiter-McLaughlin effect of WASP-69b}

The Rossiter-McLaughlin (RM) effect is an apparent radial velocity shift of the stellar spectrum during the transit of a planet. As the planet moves across the stellar disk, it occults different regions with different radial velocities due to stellar rotation. Here, we measure this effect by fitting the analytical radial-velocity model presented by \citet{2005Ohta} to the stellar radial velocity data given in the HARPS files headers, using a Markov chain Monte Carlo (MCMC) algorithm. 

The RM model depends on the scaled semi-major axis ($a/R_{\star}$), the orbital period ($P$), the central transit time ($T_0$), the ratio of the planetary and stellar radius ($R_p/R_{\star}$), the inclination of the orbit ($i_p$), the inclination of the stellar rotation axis ($i_{\star}$), the sky-projected angle between the stellar rotation axis and the normal of orbit plane ($\lambda$), the angular rotation velocity of the star ($\Omega$) and the linear limb dark coefficient ($\epsilon$). We fix $a/R_{\star}$, $P$, $T_0$, $R_p/R_{\star}$, $i_p$ and $i_{\star}$, and the rest of parameters remain free. The system was analyzed with $20$ chains with a total of ${\sim}1x10^5$ steps. Each step is started at random points near the expected values from literature and $\lambda$ is constrained to $-180-180~^o$. For WASP-69b $i_{\star}$ is fixed at $90^o$. The best-fit values are obtained at $50\%$ (median) and their error bars correspond to the $1\sigma$ statistical errors at the corresponding percentiles. 

Note that the RV information measured by HARPS is given with possible instrumental and stellar activity effects, reflected as additional offsets to the data, which vary from night to night. For this reason, we add the stellar radial velocity curve to the RM effect model, and the offsets between the data sets and the final model are fitted as free parameters, when performing an joint analysis of all available nights. The MCMC results are presented in Table~\ref{tab:RM_values} and the correlation diagrams for the probability distribution in Figure~\ref{fig:ap_W69} in the \ref{appendix}.

\begin{table*}[]
\centering
\caption{MCMC observational results of the Rossiter-McLaughlin effect for WASP-69b and HD~189733b.}
\label{tab:RM_values}
\begin{tabular}{lcc|ccc}
\hline\hline
\multirow{2}{*}{}    & \multicolumn{2}{c|}{WASP-69b}     & \multicolumn{3}{c}{HD 189733b}          \\ \cline{2-6} 
                            & \multicolumn{1}{c}{Night 1} & \multicolumn{1}{c|}{Night 2} & \multicolumn{1}{c}{Night 1} & \multicolumn{1}{c}{Night 2} & \multicolumn{1}{c}{Night 3} \\ \hline
$\chi^2$                    &  $1.1$    & $0.5$  &   $2.1$  &    $4.3$   &    $1.5$         \\
Offset $\mathrm{(km/s)}$    &  $-9.6364\pm0.0004$ &$-9.6251\pm0.0005$   &   $-2.2213^{+0.0001}_{-0.0002}$    &    $-2.2314^{+0.0001}_{-0.0002}$     &    $-2.2204\pm0.0002$              \\
$\Omega~\mathrm{(rad/day)}$ & \multicolumn{2}{c|}{$0.24^{+0.02}_{-0.01}$}   & \multicolumn{3}{c}{$0.559\pm0.003$} \\
$\lambda~\mathrm{(^o)}$     & \multicolumn{2}{c|}{$0.4^{+2.0}_{-1.9}$}     & \multicolumn{3}{c}{$-0.31\pm0.17$}               \\
$\epsilon$                  & \multicolumn{2}{c|}{$0.779\pm0.048$}      & \multicolumn{3}{c}{$0.877^{+0.011}_{-0.012}$}   \\              \hline                                        
\end{tabular}
\end{table*}

For WASP-69b we derive an stellar angular rotation of $0.24^{+0.02}_{-0.01}~\mathrm{rad/day}$, consistent with the stellar rotational period of $\approx23~\mathrm{days}$ presented in \citet{2014MNRAS.445.1114A}. The small sky-projected angle between the stellar rotation axis and the normal of the orbital plane ($\lambda = 0.4^{+2.0}_{-1.9}$), means that the sky projections of the stellar spin-axis and the orbit normal are aligned to within a few degrees. Finally, we measure a limb darkening coefficient of $0.779\pm0.048$.

The same method is used to re-measure the RM effect of HD~189733b, obtaining the results presented in Table~\ref{tab:RM_values} and Figure~\ref{fig:ap_HD189}. In this case, we fix $i_{\star}\approx92^{+12}_{-4}$ $^o$ from \citet{2016Cegla}. Previous studies retreived $\lambda$ values ranging from $-1.4\pm1.1^o$ to $-0.35\pm0.25^o$ (\citealt{2006Winn}; \citealt{2009Triaud}; \citealt{2010Cameron}; \citealt{2016Cegla}). Our result are in line with with these values. The measured angular rotation of the star ($\omega = 0.559\pm0.003~\mathrm{rad/day}$) is also consistent with the equatorial period of $11.94\pm0.16~\mathrm{days}$ presented in \citet{2010Fares}, and the resulting linear limb darkening coefficient ($\epsilon = 0.876\pm0.011$) is very similar to the one presented in \citet{2016Borsa} calculated over the HARPS wavelength range ($0.7962\pm0.0013$).


\section{Conclusions}

We present an analysis of new transit observations of WASP-69b using the HARPS-North spectrograph at Roque de los Muchachos Observatory (La Palma), and archive observations of the well-studied exoplanet HD~189733b using the HARPS-South spectrograph at La Silla (Chile). A new data analysis method is presented, which has been first applied to the HD~189733b data in order to compare with similar studies, and then to our WASP-69b observations. The method, based on correcting the telluric contamination with a one-dimensional telluric water model from \citet{Yanmodel}, can be very useful in observations where the baseline is not large enough to compute a high quality telluric spectrum as in \citet{2013A&A...557A..56A}, \citet{2015A&A...577A..62W}, \citet{2017A&A...603A..73Y}, \citet{2017A&A...602A..36W}, among others. Telluric sodium is present in WASP-69b observations, emphasizing the importance of using one of the HARPS fibers monitoring the sky simultaneously to the observations to correct for such effects. 

We observe atomic sodium (Na I) features in the transmission spectrum of both planets. For HD~189733b, the Na I doublet is clearly detected with line contrasts of $0.72\pm0.05~\%$ for the D2 and $0.51\pm0.05~\%$ for the D1, similar to those presented in \citet{2015A&A...577A..62W}, and FWHMs of $0.64\pm0.04~\mathrm{\AA}$ (D2) and $0.60\pm0.06~\mathrm{\AA}$ (D1), larger than those measured in the same paper. For WASP-69b, even if the D2 Na I line is clearly detected, only its line contrast can be measured ($5.8\pm0.3~\%$), and more transits will be needed to estimate the FHWM and reach enough SNR to characterize the D1 line. 
  
\citet{2015A&A...577A..62W} reported a blueshift of $0.16\pm0.04~\mathrm{\AA}$ in the Na I lines measured in HD~189733b's transmission spectrum. Here, a blueshift of ${\sim}0.04~\mathrm{\AA}$ ($2~\mathrm{km~s^{-1}}$) is measured, far smaller than in \citet{2008ApJ...673L..87R} (${\sim}0.75~\mathrm{\AA}$) but consistent with the the wind speed detected by \citet{2015ApJ...814L..24L} and \citet{Brogi2016}. \citet{2015A&A...577A..62W} speculate that this discrepancy can be attributed to global errors on the planetary radial-velocity considering the HARPS precision. On the other hand, no net blueshift is measured in the WASP-69b Na I lines.

The Na I transit light curves of both planets are also calculated. The observed transit light curve of HD~189733b shows the characteristic shape of the CLV effect presented in \citet{2017A&A...603A..73Y}, which is corrected in order to measure the true absorption depth at different bandwidths. For WASP-69b the CLV effect is predicted to be much lower comparing to the Na absorption, and not clearly distinguishable on the light curves, although the light curves are better fitted by a model containing this component.

By analyzing the Rossiter-McLaughlin effect of WASP-69b we obtain an stellar angular rotation velocity of $0.24^{+0.02}_{-0.01}~\mathrm{rad/day}$, consistent with the stellar rotational period presented in \citet{2014MNRAS.445.1114A} ($\approx23~\mathrm{days}$). We also measure a sky-projected angle between the stellar rotation axis and the normal of orbit plane ($\lambda$) of $0.4^{+2.0}_{-1.9}$. In the case of HD~189733b, the RM effect has been extensively studied in previous works (\citealt{2006Winn}; \citealt{2009Triaud}; \citealt{2010Cameron}; \citealt{2016Cegla}), and consistent results are obtained here.

In the near-future, high resolution spectrographs like ESPRESSO on VLT and HIRES on E-ELT will perform transit observations of many exoplanets, making it possible to study their transmission spectra with a higher signal-to-noise ratio. WASP-69b should be among the first targets of these instruments in order to further constrain sodium abundances and atmospheric temperature/pressure profiles for its atmosphere.

\begin{acknowledgements}

Based on observations made with the Italian Telescopio Nazionale Galileo (TNG) operated on the island of La Palma by the Fundación Galileo Galilei of the INAF (Istituto Nazionale di Astrofisica) at the Spanish Observatorio del Roque de los Muchachos of the Instituto de Astrofisica de Canarias. This work is partly financed by the Spanish Ministry of Economics and Competitiveness through projects ESP2014-57495-C2-1-R and ESP2016-80435-C2-2-R. We wish to thank D. Ehrenreich and A. Wyttenbach for pointing the authors to additional existing data in the HARPS archive and useful discussion. 

\end{acknowledgements}

%
%
\bibliographystyle{aa.bst} 
\bibliography{aa.bib} 





   
  



%

%
\onecolumn
\begin{appendix} 
\section{Probability distributions}
\label{appendix}

\begin{figure*}[h]
\includegraphics[width=\textwidth]{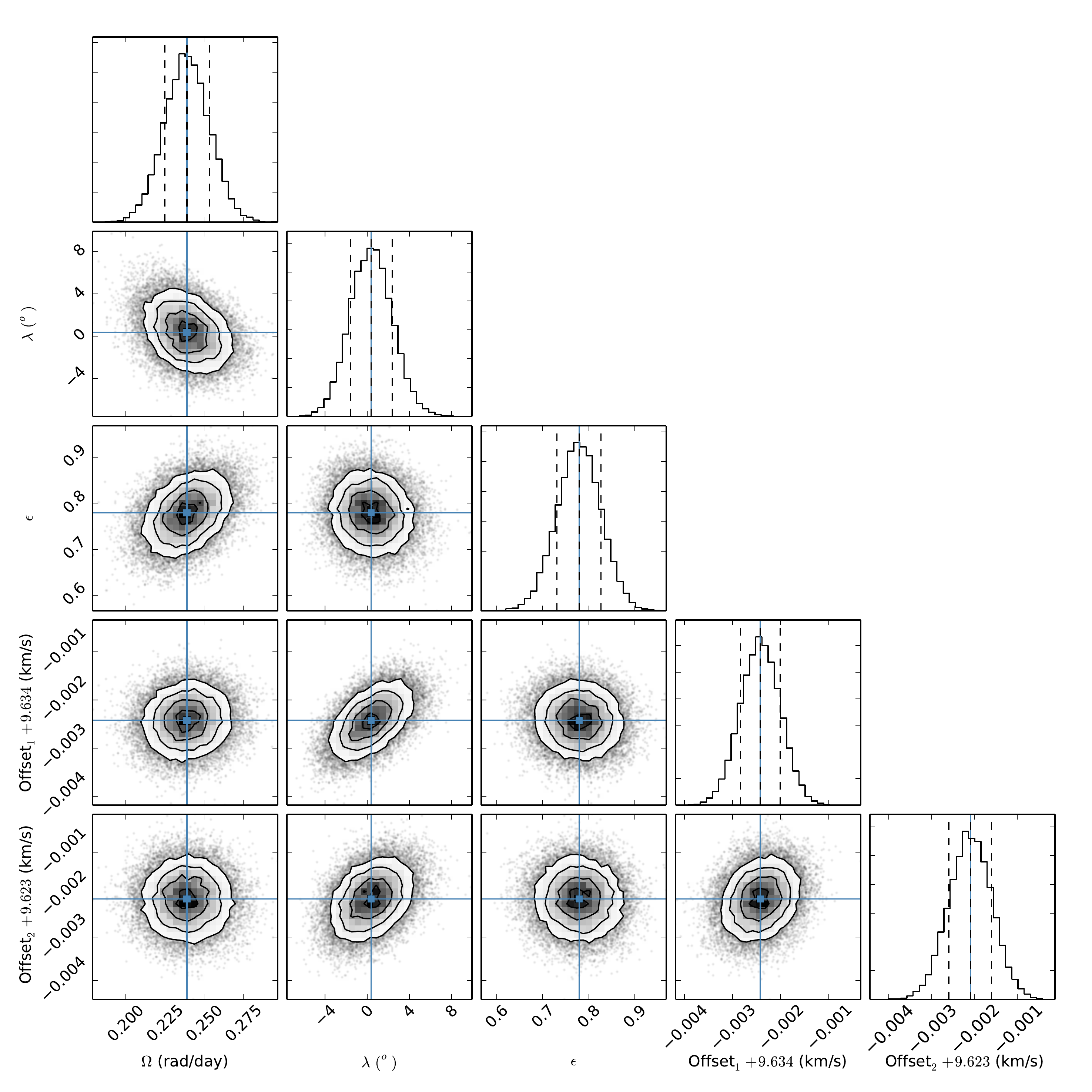}
\caption{Correlation diagrams for the probability distribution of the Rossiter-McLaughlin model parameters for WASP-69b. The dashed lines on the histograms correspond to the $16$ and $84$ percentiles used to obtain the the $1\sigma$ statistical errors. The blue lines show the resulting median values.}
\label{fig:ap_W69}
\end{figure*}

\begin{figure*}[h]
\includegraphics[width=\textwidth]{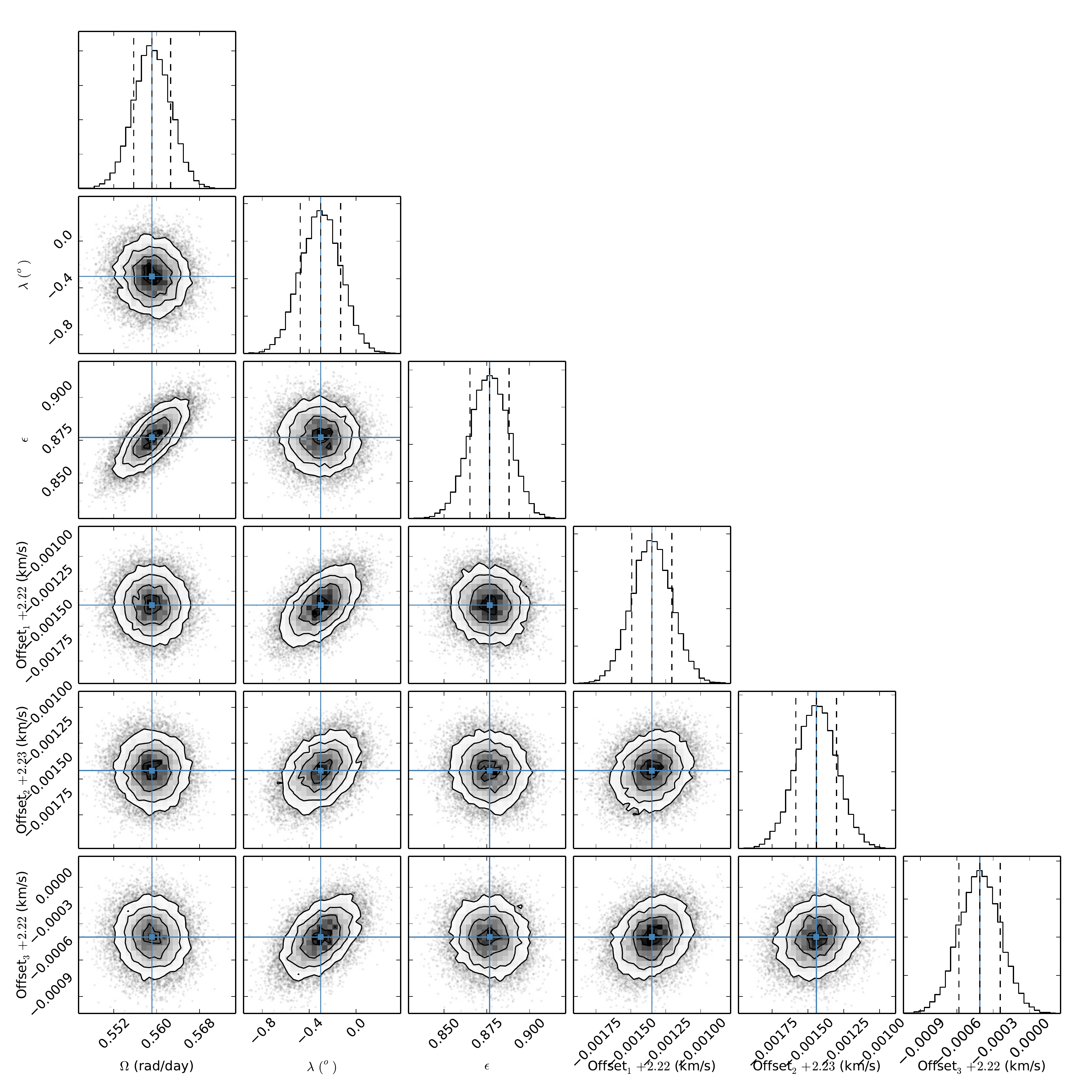}
\caption{Correlation diagrams for the probability distribution of the Rossiter-McLaughlin model parameters for HD~189733bb.}
\label{fig:ap_HD189}
\end{figure*}

\end{appendix}
\end{document}